\documentclass[11pt]{article}

\usepackage{amsfonts}
\usepackage{amsgen,amsmath,amstext,amsbsy,amsopn,amssymb}
\usepackage[dvips]{graphicx}

\usepackage[usenames]{color}

\newtheorem{Theorem}{Theorem}[section]

\newtheorem{Lemma}{Lemma}[section]
\newtheorem{Remark}{Remark}[section]

\newcommand{\be}{\begin{equation}}
\newcommand{\ee}{\end{equation}}
\newcommand{\bea}{\begin{eqnarray}}
\newcommand{\eea}{\end{eqnarray}}
\newcommand{\beas}{\begin{eqnarray*}}
\newcommand{\eeas}{\end{eqnarray*}}

\newcommand{\X}{{\mathbf{X}}}

\newcommand{\diag}{{\rm diag}}

\allowdisplaybreaks

\begin{document}

\title{Sharp RIP Bound for Sparse Signal and Low-Rank\\
  Matrix Recovery}
\author{T. Tony Cai\footnote{The research of Tony Cai was supported in part by NSF FRG Grant DMS-0854973.}~ and Anru Zhang\\
University of Pennsylvania}

\date{}
\maketitle

\begin{abstract}
This paper establishes a sharp condition on the restricted isometry property (RIP) for both the sparse signal recovery and low-rank matrix recovery. It is shown that if the measurement matrix $A$ satisfies the RIP condition  $\delta_k^A<1/3$, then all $k$-sparse signals $\beta$ can be recovered exactly via the constrained $\ell_1$ minimization based on $y=A\beta$. Similarly, if the linear map $\cal M$ satisfies the RIP condition  $\delta_r^{\cal M}<1/3$, then all matrices $X$ of rank at most $r$ can be recovered exactly via the constrained nuclear norm minimization based on $b={\cal M}(X)$. Furthermore, in both cases  it is not possible to do so in general when the condition does not hold. In addition,  noisy cases are considered and oracle inequalities are given under the sharp RIP condition.
\end{abstract}


\section{Introduction}

Compressed sensing has been a very active field of recent research with a wide range of applications, including signal processing, medical imaging, seismology, and statistics. The goal is to develop efficient data acquisition techniques that allow accurate reconstruction of highly undersampled sparse signals.  It is now well understood that the  constrained $\ell_1$ minimization method provides an effective way for recovering sparse signals. See, e.g.,  Cand\`es and Tao  \cite{Candes_Decoding, Candes_Dantzig}, Donoho \cite{Donoho06} and Donoho, Elad, and Temlyakov \cite{DET}. A closely related problem is the  affine rank minimization problem, where the goal is to recover a large low-rank matrix based on an observation of an affine transformation of the matrix. Applications include linear system identification and control, Euclidean embedding, and image compression.  See, e.g., Cand\`es and Plan \cite{Candes_Oracle}, and Recht, Fazel and Parrilo \cite{Recht_Matrix}.

More specifically, in compressed sensing, one observes $(A, y)$ with
\begin{equation}\label{eq:modelsignal}
y=A\beta+z
\end{equation}
where $y\in \mathbb{R}^n$,  $A\in \mathbb{R}^{n\times p}$ with $n\ll p$, $\beta\in \mathbb{R}^p$ is a sparse signal of interest, and $z\in\mathbb{R}^n$ is a vector of measurement errors. One wishes to recover the unknown sparse signal $\beta\in\mathbb{R}^p$ based on $A$ and $y$ using an efficient algorithm. The affine rank minimization problem aims to reconstruct a low-rank matrix $X$ based on a known linear map $\mathcal{M}$ and an observed vector $b \in \mathbb{R}^q$ where
\begin{equation}\label{eq:modelmatrix}
b=\mathcal{M}(X)+z.
\end{equation}
Here $\mathcal{M}:\mathbb{R}^{m\times n}\to \mathbb{R}^q$ is a linear map, $X\in \mathbb{R}^{m\times n}$ is an unknown low-rank matrix of interest, and $z\in\mathbb{R}^q$ is an error vector.

The methods of constrained $\ell_1$ and nuclear norm minimization,
\bea\label{eq:signalmini}
(P_\mathcal{B})&& \hat\beta=\arg\min_\beta\{ \|\beta\|_1: \; A\beta-y\in\mathcal{B}\}\\
\label{eq:matrixmini}
(P_\mathcal{B})&&  X_\ast=\arg\min_{X}\{\|X\|_\ast: \;  \mathcal{M}(X)-b\in\mathcal{B}\},
\eea
as  convex relaxations to $\ell_0$ and rank minimization respectively, have been shown to be very effective in solving these problems. Here  $\|\X\|_\ast$ is the nuclear norm of $X$, which is defined to be the sum of the singular values of $X$, and $\mathcal{B}$ is a bounded set determined by the noise structure. For example, $\mathcal{B} = \{0\}$ in the noiseless case and $\mathcal{B}$ is the feasible set of the error vector $z$ in the case of bounded noise.

One of the most commonly used frameworks for sparse signal and low-rank matrix recovery is the \emph{Restricted Isometry Property} (RIP). See Cand\`es and Tao \cite{Candes_Decoding} and Recht et al. \cite{Recht_Matrix}. A vector is  said to be $k$-sparse if $|{\rm supp}(v)|\leq k$, where ${\rm supp}(v)=\{i:v_i\neq 0\}$ is the support of $v$. In this paper, we shall use the phrase``$r$-rank matrices" to refer to matrices of rank at most $r$.
In compressed sensing, the RIP  requires subsets of certain cardinality of the columns of $A$ to be close to an orthonormal system. The RIP conditions for the signal and matrix recovery are similar and we shall state them together to save space.
Let $A\in \mathbb{R}^{n\times p}$ be a matrix and let $\mathcal{M}: \mathbb{R}^{m\times n}\to \mathbb{R}^q$ be a linear map. For integers $1\leq k\leq p$ and $1\leq r\leq \min\{m,n\}$, define the restricted isometry constants (RIC) $\delta_k^A$ and $\delta_r^{\mathcal{M}}$ to be the smallest non-negative numbers  such that for all $k$-sparse vectors $\beta$ and all $r$-rank matrices $X$,
\bea
(1-\delta_k^{A})\|\beta\|^2_2\; \leq &\|A \beta\|^2_2& \leq \; (1+\delta_k^{A})\|\beta\|_2^2 \label{eq:ripsignal}\\
(1-\delta_r^{\mathcal{M}})\|X\|^2_F\; \leq & \|\mathcal{M}(X)\|_2^2 & \leq \; (1+\delta_r^{\mathcal{M}})\|X\|_F^2
\label{eq:ripmatrix}
\eea
where $\|X\|_F^2=\sum x_{ij}^2$ is the squared Frobenius norm of $X=(x_{ij})$.

A major focus of compressed sensing is to find explicit and simple conditions under which the sparse signals can be recovered exactly using a computational efficient algorithm. A variety of sufficient conditions on the RIP  for the exact/stable recovery of $k$-sparse signals and $r$-rank matrices have been introduced in the literature. Sufficient conditions for the signal recovery include  $\delta_{2k}^A<\sqrt{2}-1$ in Cand\`es \cite{Candes08},  $\delta_{2k}^A<0.472$ in Cai, Wang and Xu \cite{Cai_Shift}, $\delta_{2k}<0.493$ in Mo and Li \cite{Mo} and $\delta_k^A<0.307$ in Cai, Wang and Xu \cite{Cai_Newbound}; for the matrix recovery, sufficient conditions are $\delta_{4r}^\mathcal{M}<\sqrt{2}-1$ in Cand\`es and Plan \cite{Candes_Oracle}, $\delta_{5r}^\mathcal{M}<0.607$, $\delta_{4r}^\mathcal{M}<0.558$, $\delta_{3r}^\mathcal{M}<0.4721$ in Mohan and Fazel \cite{Mohan}, $\delta_{2r}^{\mathcal{M}}<0.4931$, $\delta_{r}^\mathcal{M}<0.307$ in Wang and Li \cite{Wang_NewRIC}.  On the other hands, negative results have also been obtained. In the case of signal recovery, Davies and Gribonval \cite{Davies_RIPfail}  and Cai, Wang and Xu \cite{Cai_Newbound} showed
respectively that it is impossible to recover certain $k$-sparse signals when $\delta_{2k}^A> \sqrt{2}/2$ and when $\delta_k^A={k-1\over 2k-1}<0.5$. For matrix recovery, Wang and Li \cite{Wang_NewRIC}  proved that nuclear norm minimization cannot recover exactly all rank $r$ matrices  in the noiseless case when $\delta_r^\mathcal{M}=1/3$ or $\delta_{2r}^\mathcal{M} = \sqrt{2}/2 +\varepsilon$, where $\varepsilon$ is arbitrarily small.

Among those RIP conditions,  the ones on $\delta_{k}^A$ and $\delta_{r}^\mathcal{M}$ are arguably the most natural for the reconstruction of $k$-sparse signals and $r$-rank matrices, respectively. The main goal of this paper is to establish a sharp condition on $\delta_k^A$ and $\delta_r^\mathcal{M}$. Specifically, we show that in the noiseless case ($z=0$) the conditions
\be
\label{RIP}
\delta_k^A<{1\over 3} \quad\mbox{and} \quad \delta_k^\mathcal{M}<{1\over 3}
\ee
are sharp respectively for the exact recovery of $k$-sparse signals based on \eqref{eq:modelsignal} and for the exact recovery of $r$-rank matrices based on \eqref{eq:modelmatrix}. These conditions are also sharp for the stable recovery in the noisy case. That is, under the condition  $\delta_k^A<1/3$, all $k$-sparse signals can be exactly recovered via the constrained $\ell_1$ minimization \eqref{eq:signalmini} in the noiseless case and can be stably recovered in the noisy case. Furthermore, it is not possible to do so in general if $\delta_k^A\geq1/3$.
Similarly, for the recovery of $r$-rank matrices using the constrained nuclear norm minimization based on  \eqref{eq:modelmatrix}, the condition  $\delta_r^\mathcal{M}<{1/3}$ is sharp. To the best of our knowledge, \eqref{RIP} is the first sharp RIP condition.

Various oracle inequalities have been given in the literature for the constrained $\ell_1$/nuclear norm minimization estimators, known as the Dantzig Selector, in the setting of Gaussian noise. See, for example, Cand\`es and Tao \cite{Candes_Dantzig} and Cai, Wang and Xu \cite{Cai_Stable} for the sparse signal recovery and  Cand\`es and Plan  \cite{Candes_Oracle} for the matrix recovery under the condition $\delta_{4r}^\mathcal{M}<\sqrt{2}-1$. In this paper we derive oracle inequalities for both sparse signal and low-rank matrix recovery  under the condition $\delta_k^A<1/3$ and $\delta_r^\mathcal{M}<1/3$.

Besides providing a sharp condition on $\delta_{k}^A$ and $\delta_{r}^\mathcal{M}$, the same techniques can also be used to sharpen other RIP conditions such as $\delta_{2k}^A$ and $\delta_{2r}^\mathcal{M}$. We show that, in the noiseless case,  $\delta_{2k}^A\le 1/2$ and $\delta_{2r}^\mathcal{M} \le 1/2$ are respectively sufficient for the exact recovery of $k$-sparse signals based on \eqref{eq:modelsignal} and for the exact recovery of $r$-rank matrices based on  \eqref{eq:modelmatrix}.

The rest of the paper is organized as follows. Section \ref{notation.sec} reviews basic notations and
definitions and states some useful facts on the null spaces. Section \ref{division.sec} then introduces a technically important tool called the Division Lemma, which is used in the detailed analysis for both the signal and matrix recovery.  Sections \ref{signal.sec} and \ref{matrix.sec} separately analyze the sparse signal recovery and low-rank matrix recovery, in both the noiseless and noisy settings.  Section \ref{discussion.sec} provides oracle inequalities for Gaussian noise under the conditions $\delta_k^A<1/3$ and $\delta_r^\mathcal{M}<{1/3}$, and discusses other RIP conditions. The proofs of the main results are given in Section \ref{proofs.sec}.

\section{Notations and Preliminaries}
\label{notation.sec}

In this section, we introduce basic notations and definitions that will be used throughout the paper, and state some facts on the null spaces that will be used later.

 For a vector $v=(v_1,\cdots,v_p)'\in \mathbb{R}^p$, define $v_{\max(k)}$ to be the vector $v$ with all but the largest $k$ entries in absolute values set to zero, and let $v_{-\max(k)}=v-v_{\max(k)}$. For a matrix $X\in \mathbb{R}^{m\times n}$ (without loss of generality, assume that $m\leq n$), let $a_1\ge a_2 \ge \cdots \ge a_m$ be its singular values and let $X=\sum_{i=1}^ma_iu_iv_i^T$ be the singular value decomposition of $X$. We define $X_{\max(r)}=\sum_{i=1}^ra_iu_iv_i^T$ and $X_{-\max(r)}=X-X_{\max(r)}=\sum_{i=r+1}^m a_iu_iv_i^T$.

For $0<\alpha<\infty$ define the $\ell_\alpha$ norm of a vector $v\in\mathbb{R}^p$ as $\|v\|_\alpha=(\sum_{i=1}^p|v_i|^\alpha)^{1/\alpha}$. In addition, $\|v\|_\infty=\sup_i|v_i|$ and $\|v\|_0=|{\rm supp}(v)|$. For matrices $X=(x_{ij}),\, Y=(y_{ij})\in \mathbb{R}^{m\times n}$, define the inner product of $X$ and $Y$ as $\langle X, Y\rangle={\rm trace}(X^T Y)=\sum_{i=1}^m\sum_{j=1}^n x_{ij}y_{ij}$. The norm associated with this inner product is the Frobenius norm, $\|X\|_F=\sqrt{\langle X,X\rangle}=\sqrt{\sum_{i=1}^m\sum_{j=1}^nx_{ij}^2}$.
Note that $\mathbb{R}^{m\times n}$ associated with this inner product is a Hilbert space. The spectral norm of a matrix $X\in \mathbb{R}^{m\times n}$ is defined as $\|X\|=\sup_{\beta\in\mathbb{R}^n}\|X\beta\|_2/\|\beta\|_2$, which is equal to the largest singular value of $X$.

For a linear map  $\mathcal{M}:\mathbb{R}^{m\times n}\to\mathbb{R}^q$, we denote its  adjoint operator by $\mathcal{M}^*: \mathbb{R}^q\to\mathbb{R}^{m\times n}$, so that for all $X\in \mathbb{R}^{m\times n}$ and $b\in \mathbb{R}^q$,  $\langle X,\mathcal{M}^\ast(b) \rangle=\langle\mathcal{M}(X),b\rangle_{\ell_2}$.
For any given norm $|\cdot|$ in an inner product space $(\mathbb{R}^{m\times n},\langle\cdot,\cdot\rangle)$,  the dual norm $|\cdot|_d$ is defined as $|X|_d=\max\{\langle X,Y\rangle:|Y|=1\}$. It is well known that the dual norm of the Frobenius norm is itself and the nuclear norm and spectral norm are dual norms of each other.
The null spaces of a matrix $A\in\mathbb{R}^{n\times p}$ and a linear map $\mathcal{M}:\mathbb{R}^{m\times n}\to\mathbb{R}^q$ are denoted respectively by $\mathcal{N}(A)$ and $\mathcal{N}(\mathcal{M})$, i. e.,
$\mathcal{N}(A)=\{\beta\in\mathbb{R}^p:A\beta=0\}$ and $\mathcal{N}(\mathcal{M})=\{X\in\mathbb{R}^{m\times n}:\mathcal{M}(X)=0\}$.

Finally, we introduce a useful tool for providing conditions for the exact recovery. Stojnic, Xu, Hassibi \cite{Stojnic} gave a necessary and sufficient  condition on the null space for the exact recovery of $k$-sparse signals in the noiseless case. It was shown that one can recover all $k$-sparse signals $\beta$ using \eqref{eq:signalmini} with $\mathcal{B}=\{0\}$  if and only if for all $\beta\in\mathcal{N}(A)\setminus \{0\}$,
$$\|\beta_{\max(k)}\|_1<\|\beta_{-\max(k)}\|_1.\eqno{(\ast)}$$
Oymak, Hassibi \cite{Oymak} gave a similar result for the exact recovery of $r$-rank matrices in the noiseless case. One can recover all $r$-rank matrices $X$ using \eqref{eq:matrixmini} with $\mathcal{B}=\{0\}$ if and only if for all $X\in\mathcal{N}(\mathcal{M})\setminus \{0\}$,
$$\|X_{\max(r)}\|_\ast<\|X_{-\max(r)}\|_\ast.\eqno{(\ast\ast)}$$
Based on these results, one can consider the recovery problem by investigating the null spaces of $A$ and  $\mathcal{M}$ instead of checking the original definition of exact recovery, which often simplifies the problem.

\section{Sharp RIP conditions for Sparse Signal and Low-rank Matrix Recovery}
\label{optimal.sec}

With the preparations given in Section \ref{notation.sec}, we establish in this section the main results of this paper -- a sharp RIP bound for the exact recovery of  sparse signals and low-rank matrices in the noiseless case and the stable recovery in the noisy case. A unified treatment is given for the sparse signal recovery and low-rank matrix recovery.  We first introduce in Section \ref{division.sec} an elementary but important technical lemma which we call the \emph{Division Lemma}, and then discuss the main results for sparse signal recovery in Section \ref{signal.sec} and the low-rank matrix recovery in Section \ref{matrix.sec}. 

\subsection{Division Lemma}
\label{division.sec}

As discussed in Section \ref{notation.sec}, we will establish the RIP condition for the exact recovery using the null space properties of $A$ and $\mathcal{M}$. In order to relate the general elements in the null space with the RIP condition whose constraint is on the sparse vectors and low-rank matrices, a natural approach is to divide these elements into sums of sparse or low-rank components. Consequently, we introduce the Division Lemma below, which is a key technical tool for the proof of the main results.
\begin{Lemma}[Division Lemma] \label{lm:divide}
Let $r$ and $m$ be positive integers with $m\geq 2r $. Let $a_1\geq a_2\geq a_3\geq\cdots\geq a_m\geq 0$ be a sequence of non-increasing real numbers satisfying
\begin{equation}\label{eq:a1>=am}
\sum_{w=1}^r a_w \geq \sum_{w=r+1}^m a_w.
\end{equation}
Then there exist non-negative real numbers $\{s_{ij}\}_{1\leq i\leq r, 2r+1\leq j\leq m}$ such that
\begin{equation}\label{eq:divide1}
\sum_{i=1}^rs_{ij}=a_j,\quad \forall \; 2r+1\leq j\leq m,
\end{equation}
and
\begin{equation}\label{eq:divide2}
\frac{1}{r}\sum_{w=1}^ra_{w}\geq a_{r+i}+\sum_{j=2r+1}^m s_{ij} ,\quad \forall \; 1\leq i\leq r.
\end{equation}
\end{Lemma}

The proof of Lemma \ref{lm:divide} is simply by induction on $m$. The Division Lemma can be illustrated as in the following table. Each row is an inequality; every element in the first row equals the sum of remaining elements in the same column:

\begin{center}
\begin{tabular}{cccccc|cccccccc}
  \hline
  $a_1$ & $a_2$  &$\cdots$& $a_r$  &   & $\geq$ & $a_{r+1}$ & $a_{r+2}$ &$\cdots$ & $a_{2r}$& $+$ & $a_{2r+1}$ & $\cdots$ & $a_m$ \\\hline
$a_1/r$ &$a_2/r$ &$\cdots$&$a_r/r$ &  & $\geq$ & $a_{r+1}$ &           &         &         & $+$ & $s_{1,2r+1}$ & $\cdots$ & $s_{1,m}$ \\
$a_1/r$ &$a_2/r$ &$\cdots$&$a_r/r$ &  & $\geq$ &           & $a_{r+2}$ &         &         & $+$ & $s_{2,2r+1}$ & $\cdots$ & $s_{2,m}$ \\
$\vdots$&$\vdots$&$\ddots$&$\vdots$&  & $\geq$ &           &           &$\ddots$ &         & $+$ & $\vdots$ &  & $\vdots$ \\
$a_1/r$ &$a_2/r$ &$\cdots$&$a_r/r$ &  & $\geq$ &           &           &         & $a_{2r}$& $+$ & $s_{r,2r+1}$ & $\cdots$ & $s_{r,m}$ \\
  \hline
\end{tabular}
\end{center}

\subsection{Sparse signal recovery}
\label{signal.sec}

We begin with the noiseless case ($z=0$) of the sparse signal recovery model \eqref{eq:modelsignal}. In this case, The commonly used $\ell_1$ minimization method is by \eqref{eq:signalmini} with $\mathcal{B}=\{0\}$. We shall present the sharp RIP condition on  $\delta_k^A$ for the exact recovery of all $k$-sparse signals  for any given integer $k\ge 2$.

The following theorem shows that the condition $\delta_{k}^A<1/3$ is sufficient for the exact recovery of  $k$-sparse signals in the noiseless case.
\begin{Theorem}\label{th:mainsignal}
 Suppose the measurement matrix $A\in \mathbb{R}^{n\times p}$ satisfies $\delta_{k}^A<1/3$ for some integer $2\le k\le p$. Let $y = A \beta$ where $\beta\in\mathbb{R}^p$ is a $k$-sparse vector.  Then the minimizer $\hat\beta$ of \eqref{eq:signalmini} with $\mathcal{B}=\{0\}$ recovers $\beta$ exactly, i.e., $\hat\beta=\beta$.
\end{Theorem}

The result below shows that the condition $\delta_k^A<1/3$ is sharp for the exact recovery in the noiseless case.

\begin{Theorem}\label{th:counterexamplesignal}
Let $2\leq k\le p/2$. There exists a measurement matrix $A\in\mathbb{R}^{n\times p}$ with $\delta_k^{A}= 1/3$ such that
for some $k$-sparse signals $\gamma, \, \eta\in\mathbb{R}^{p}$ with $\gamma\neq \eta$,  $A\gamma = A\eta$. Consequently, it is not possible for any method to exactly recover all $k$-sparse signals $\beta$ based on $(A, y)$ with $y=A\beta$.
In particular, the $\ell_1$ minimization (\ref{eq:signalmini}) with $\mathcal{B}=\{0\}$ cannot recover all $k$-sparse signals. 
\end{Theorem}

Theorems \ref{th:mainsignal} and \ref{th:counterexamplesignal} together show that the condition $\delta_k^{A}< 1/3$ is sharp for all $2\leq k\le p/2$.

\begin{Remark}{\rm
In the above theorems, The case $k=1$ is excluded because the RIP cannot provide any sufficient condition for the exact recovery via the constrained $\ell_1$ minimization in this case. Take, for example, $n=p-1\ge 1$. Let $A\in \mathbb{R}^{n\times p}$ with $A\beta = (\beta_1-\beta_2,\beta_3,\beta_4,\cdots,\beta_p)^T$ for any $\beta=( \beta_1, \beta_2,\beta_3,\cdots,\beta_p)^T\in \mathbb{R}^{p}$.
Then for all 1-sparse vectors $\beta$,
$$\|A \beta\|_2^2=\sum_{i=1}^p\beta_i^2-2\beta_1\beta_2=\|\beta\|_2^2,$$
which implies the restricted isometry constant $\delta_1^A=0$. However, $b= A\gamma=A\eta$ where $\gamma=(1,0,\cdots,0)$ and $\eta=(0,-1,0,\cdots,0)$ are both 1-sparse signals. Thus it is impossible to recover both of them exactly relying only on the information of $(A, b)$. In particular,  the $\ell_1$  minimization (\ref{eq:signalmini}) with $\mathcal{B}=\{0\}$ cannot recover all $1$-sparse signals.  Since $\delta_1^A=0$, the RIP cannot provide any sufficient condition in this case.
}
\end{Remark}

We shall now turn to the noisy case of the sparse signal recovery model \eqref{eq:modelsignal}. The noiseless case provides much insight to the noisy case. In this case the error vector $z$ is nonzero and we shall consider two bounded noise settings
\bea
\mathcal{B}^{\ell_2}(\eta)&=&\{z:\|z\|_2\leq\eta\}, \label{eq:B^l2signal}\\
\mathcal{B}^{DS}(\eta)&=&\{z:\|A^Tz\|_\infty\leq\eta\}. \label{eq:B^DSsignal}
\eea
The case of Gaussian noise, which is a canonical model in statistics, can be treated similarly. See Remark \ref{Gaussian.remark} below.
In the noisy case we shall also consider more general signals $\beta$ which are not necessarily $k$-sparse. Decompose $\beta=\beta_{\max(k)}+\beta_{-\max(k)}$.   The $\ell_1$ norm minimization approach for recovering $\beta$ in these bounded noise settings is by solving \eqref{eq:signalmini} with $\mathcal{B}=\mathcal{B}^{\ell_2}(\eta)$ or $\mathcal{B}=\mathcal{B}^{DS}(\eta)$.

We first consider the stable recovery of $\beta$ with the error $z $ in a bounded $\ell_2$ ball.
\begin{Theorem}\label{th:noisysignal}
Consider the signal recovery model \eqref{eq:modelsignal} with $\|z\|_2\leq\varepsilon$. Let $\hat{\beta}$ be the minimizer of \eqref{eq:signalmini} with $\mathcal{B}=\mathcal{B}^{\ell_2}(\eta)$ defined in \eqref{eq:B^l2signal} for some $\eta\ge \epsilon$. If $\delta=\delta_{k}^{A}<1/3$ with $k\geq 2$, then
\begin{equation}
\|\hat\beta-\beta\|_2\leq\frac{\sqrt{2(1+\delta)}}{1-3\delta}(\varepsilon+\eta)+\frac{2\sqrt{2}(2\delta+\sqrt{(1-3\delta)\delta})+2(1-3\delta)}{1-3\delta}\frac{\|\beta_{-\max(k)}\|_1}{\sqrt{k}}.
\end{equation}
In particular, for all $k$-sparse signals $\beta$,
\[
\|\hat\beta-\beta\|_2\leq\frac{\sqrt{2(1+\delta)}}{1-3\delta}(\varepsilon+\eta).
\]
\end{Theorem}

The result is similar if  the error $z$ is in the bounded set $\|A^Tz\|_\infty\leq\varepsilon$. The $\ell_1$ minimization method with $\mathcal{B}=\mathcal{B}^{DS}$ is called the Dantzig Selector. See Cand\`es and Tao \cite{Candes_Dantzig}.
\begin{Theorem}\label{th:noisydantzigsignal}
Consider the signal recovery model \eqref{eq:modelsignal} with $\|A^Tz\|_\infty\leq\varepsilon$. Let $\hat\beta$ be the minimizer of \eqref{eq:signalmini} with $\mathcal{B}=\mathcal{B}^{DS}(\eta)$ defined in \eqref{eq:B^DSsignal} for some $\eta\ge \epsilon$. If $\delta=\delta_{k}^{A}<1/3$ with $k\geq 2$,  then
\begin{equation}
\|\hat\beta-\beta\|_2\leq\frac{\sqrt{2k}}{1-3\delta}(\varepsilon+\eta)+\frac{2\sqrt{2}(2\delta+\sqrt{(1-3\delta)\delta})+2(1-3\delta)}{1-3\delta}\frac{\|\beta_{-\max(k)}\|_1}{\sqrt{k}}.
\end{equation}
\end{Theorem}

\begin{Remark}\label{Gaussian.remark}
{\rm
 Since Gaussian noise is essentially bounded, the results for the signal recovery in Theorems \ref{th:noisysignal} and \ref{th:noisydantzigsignal} can be directly applied to the Gaussian noise case. Interested readers are referred to Section 4 in \cite{Cai_Shift} and Lemma 5.1 in \cite{Cai_l1} for details.
}
\end{Remark}

\subsection{Low-rank matrix recovery}
\label{matrix.sec}

We now turn to the affine rank minimization problem. As mentioned before, the results are parallel to those for the sparse signal recovery. As in Section \ref{signal.sec},  we begin with the noiseless case. The ideas and results can be extended to the noisy case later. Consider the matrix recovery model \eqref{eq:modelmatrix} with $z=0$. The nuclear norm minimization method in this case is given by \eqref{eq:matrixmini} with $\mathcal{B}=\{0\}$. The goal is to recover the matrix $X$ whose rank is at most $r$.

For the same reason as in the signal recovery problem, we shall only consider the case  $r\geq2$. The following two theorems, which are parallel to Theorems \ref{th:mainsignal} and  \ref{th:counterexamplesignal},  are the main results in this paper for the low-rank matrix recovery. Theorem \ref{th:main} shows that the condition $\delta_{r}^{\mathcal{M}}<1/3$ is sufficient for the exact recovery of $r$-rank matrices.

\begin{Theorem}\label{th:main}
Suppose $2\leq r\leq \min(m,n)$. Let $X$ be a matrix of rank at most $r$ and let $b=\mathcal{M}(X)$. If $\delta_{r}^{\mathcal{M}}<1/3$, then the solution $X_\ast$  of the nuclear norm minimization \eqref{eq:matrixmini} with $\mathcal{B}=\{0\}$ recovers $X$ exactly, i.e., $X_\ast=X$.
\end{Theorem}

The following theorem shows that the condition $\delta_{r}^{\mathcal{M}}<1/3$ is sharp. These results together establish the optimal bound on $\delta_r^\mathcal{M}$ for the exact recovery in the noiseless case.
\begin{Theorem}\label{th:counterexample}
Let $2\le r \le \min(m,n)/2$. there exists a linear map $\mathcal{M}$ with $\delta_r^{\mathcal{M}}= 1/3$ such that for some matrices $X, \, Y\in\mathbb{R}^{m\times n}$ with ${\rm rank}(X), \; {\rm rank}(Y)\le r$, $\mathcal{M}(X)=\mathcal{M}(Y)$. Consequently, there does not exist any method that can exactly recover all matrices of rank at most $r$ based on $(\mathcal{M}, b)$ with $b=\mathcal{M}(X)$. In particular, the nuclear norm minimization (\ref{eq:matrixmini}) with $\mathcal{B}=\{0\}$ cannot recover all $r$-rank matrices.
\end{Theorem}

We should note that the result above is stronger than Theorem 1.2 in Wang and Li  \cite{Wang_NewRIC} as it shows that there exists some linear map $\mathcal{M}$ with $\delta_r^\mathcal{M}=1/3$ such that all methods, not just nuclear norm minimization, fail to recover all  rank $r$ matrices in the noiseless case.

\begin{Remark}{\rm
The reason for excluding the case  $r=1$ in the two theorems given above is the same as that in the signal recovery problem: the RIP cannot provide any sufficient condition in this case for the exact recovery through the nuclear norm minimization. An example is given as follows. Let $m,n\geq2$ and let the linear map $\mathcal{M}: \mathbb{R}^{m\times n} \to \mathbb{R}^{mn-2}$ be defined by
\[
\mathcal{M}(X)=(x_{11}-x_{22},x_{12}+x_{21},x_{13},\cdots,x_{1n},x_{23}\cdots,x_{2n},x_{31},\cdots,x_{mn})^T
\]
for $X=(x_{ij}) \in \mathbb{R}^{m\times n}$.
Then for all matrices $X$ such that ${\rm rank}(X)\leq 1$,
$$\|\mathcal{M}(X)\|_2^2=\sum_{i=1}^m\sum_{j=1}^nx_{ij}^2-2(x_{11}x_{22}-x_{12}x_{21})=\|X\|_F^2.$$
This implies the restricted isometry constant $\delta_r^\mathcal{M}=0$. In addition, one can check that $X=\diag(1,0,\cdots,0),\; Y=\diag(0,-1,0,\cdots,0)$ are both of rank 1. In addition, $b=\mathcal{M}(X)=\mathcal{M}(Y)$. This means that the exact recovery is impossible based on $(\mathcal{M}, b)$ in the noiseless case.
 Hence the RIP cannot provide a sufficient condition to ensure the exact recovery of all  matrices with rank at most $1$.
}
\end{Remark}

We now turn to the noisy case. As in the signal recovery problem, we also consider bounded noise in two settings
\bea
\mathcal{B}^{\ell_2}(\eta) & =& \{z:\|z\|_2\leq\eta\},\label{eq:B^l2}\\
\mathcal{B}^{DS}(\eta) &=&\{z:\|\mathcal{M}^*z\|\leq\eta\} \label{eq:B^DS}.
\eea
We shall also consider general matrices that are not necessarily exactly  low-rank. Decompose $X=X_{\max(r)}+X_{-\max(r)}$. The nuclear norm minimization method is to recover $X$ by solving \eqref{eq:matrixmini} with $\mathcal{B}=\mathcal{B}^{\ell_2}(\eta)$ or $\mathcal{B}=\mathcal{B}^{DS}(\eta)$.

We first consider the case where the error  $z$ is in a bounded $\ell_2$ ball, $\|z\|_2\leq \varepsilon$.
\begin{Theorem}\label{th:noisy}
Consider the affine rank minimization problem \eqref{eq:modelmatrix} with $ \|z\|_2\leq\varepsilon$. Let $X_\ast$ be the minimizer of \eqref{eq:matrixmini} with $\mathcal{B}=\mathcal{B}^{\ell_2}(\eta)$ defined in \eqref{eq:B^l2} for some $\eta\ge \epsilon$. If $\delta_{r}^{\mathcal{M}}<1/3$ with $r\geq 2$, then
\begin{equation}
\|X_\ast-X\|_F\leq\frac{\sqrt{2(1+\delta)}}{1-3\delta}(\varepsilon+\eta)+\frac{2\sqrt{2}(2\delta+\sqrt{(1-3\delta)\delta})+2(1-3\delta)}{1-3\delta}\frac{\|X_{-\max(r)}\|_\ast}{\sqrt{r}}.
\end{equation}
\end{Theorem}

For matrix recovery under the model \eqref{eq:modelmatrix} with the error bound $\|\mathcal{M}^\ast(z)\|\leq\varepsilon$, we have the following similar result for the matrix Dantzig Selector.
\begin{Theorem}\label{th:noisydantzig}
Consider the affine rank minimization problem \eqref{eq:modelmatrix} with $\|\mathcal{M}^\ast(z)\|\leq\varepsilon$. Let $X_\ast$ be the minimizer of \eqref{eq:matrixmini} with $\mathcal{B}=\mathcal{B}^{DS}(\eta)$ defined in \eqref{eq:B^DS} for some $\eta\ge \epsilon$. If $\delta_{r}^{\mathcal{M}}<1/3$ with $r\geq 2$, then
\begin{equation}
\|X_\ast-X\|_F\leq\frac{\sqrt{2r}}{1-3\delta}(\varepsilon+\eta)+\frac{2\sqrt{2}(2\delta+\sqrt{(1-3\delta)\delta})+2(1-3\delta)}{1-3\delta}\frac{\|X_{-\max(k)}\|_\ast}{\sqrt{r}}.
\end{equation}
\end{Theorem}
We omit the proof of Theorem \ref{th:noisydantzig}, which is essentially the same as that of Theorem \ref{th:noisy}.

\begin{Remark}{\rm
Similarly as in the sparse signal recovery problem, the results for the low-rank matrix recovery in Theorems \ref{th:noisy} and \ref{th:noisydantzig} can be extended to the Gaussian noise case. The readers are referred to Lemma 1.1 in Cand\`es and Plan \cite{Candes_Oracle} for details.
}
\end{Remark}

\section{Oracle inequalities and RIP conditions on $\delta_{2k}^A$ and $\delta_{2r}^\mathcal{M}$}
\label{discussion.sec}

Oracle inequality provides great insight into the performance of a procedure as compared to that of an ideal estimator. It was first introduced in Donoho and Johnstone \cite{DJ} in the context of statistical signal processing using wavelet thresholding. This method has since been applied in many other problems. In particular,
various oracle inequalities have been given in the literature for the constrained $\ell_1$/nuclear norm minimization procedures. See, for example, Cand\`es and Tao \cite{Candes_Dantzig}, Cai, Wang and Xu \cite{Cai_Stable}, and
Cand\`es and Plan  \cite{Candes_Oracle}. Theorem \ref{th:oracle} below provides oracle inequalities for sparse signal and low-rank matrix recovery  under the condition $\delta_k^A<1/3$ and $\delta_r^\mathcal{M}<1/3$ given in this paper. The technique is analogous to the one used in Cand\`es and Plan \cite{Candes_Oracle}, along with Lemma \ref{lm:delta2k<3deltak} given below, Theorem \ref{th:noisydantzigsignal} and Theorem \ref{th:noisydantzig}.
\begin{Theorem}\label{th:oracle}
Given the signal recovery model \eqref{eq:modelsignal}, suppose  $z\sim N_p(0, \sigma^2I)$ and the signal $\beta\in\mathbb{R}^p$ is $k$-sparse. Assume that $\hat\beta$ is the minimizer of  \eqref{eq:signalmini} with $\mathcal{B}=\{z:\|A^Tz\|_\infty\leq \lambda=4\sigma\sqrt{(2/3)\log p}\}$.
If $\delta_k^A<1/3$ with $k\geq 2$, then
\begin{equation}\label{eq:signaloracle}
\|\hat\beta-\beta\|_2^2\leq \frac{512}{3(1-3\delta_k^A)^2}\log p\sum_i\min(\beta_i^2,\sigma^2)
\end{equation}
with probability at least $1-\frac{1}{\sqrt{\pi\log p}}$.

Similarly, for the matrix case \eqref{eq:modelmatrix}, suppose $z\sim N_q(0,\sigma^2I)$ and $rank(X)\leq r$. Assume that $X_\ast$ is the minimizer of  \eqref{eq:matrixmini} with $\mathcal{B}=\{z:\|\mathcal{M}^\ast(z)\|\leq\lambda=16\sigma\sqrt{(1/3)\log(12)\max(m,n)}\}$.
 If $\delta_r^\mathcal{M}<1/3$ with $r\geq2$, then
\begin{equation}\label{eq:matrixoracle}
\|X_\ast-X\|_F^2\leq \frac{2^{12}\log12}{3(1-3\delta_r^\mathcal{M})^2}\sum_i\min(\sigma_i^2(X),\max(m,n)\sigma^2)
\end{equation}
with probability at least $1-e^{-c\max(m,n)}$, where $c>0$ is an absolute constant, and $\sigma_i(X)$, $i=1\cdots,\min(m,n)$ are the singular values of $X$.
\end{Theorem}
We should note that the main ideas for the proof here are essentially the same as those for the proof of Theorem 2.6 in \cite{Candes_Oracle}, where readers can find more details. Finally, it is noteworthy from these oracle inequalities that in the case of $\beta=0$ or $X=0$, i.e., the input signal or matrix is identically zero, the Dantzig Selector recovers the zero input exactly by  zero with high probability in the Gaussian noise case.

In addition to providing the sharp condition on $\delta_k^A$ and $\delta_r^\mathcal{M}$, the techniques developed in this paper can also be applied to sharpen other RIP conditions  such as $\delta_{2k}^A$ and $\delta_{2r}^\mathcal{M}$ for the exact/stable recovery of the sparse signals and low-rank matrices. Since $\delta_{2k}<1$ is known as a necessary condition for the model identifiability (see Lemma 1.2 in \cite{Candes_Decoding}), much previous attention has been on the bounds for $\delta_{2k}^A$ and $\delta_{2r}^\mathcal{M}$ as the sufficient conditions for the recovery of  the sparse signals and low-rank matrices.  Applying the same method as that used in the previous section on $\delta_r^\mathcal{M}$ and $\delta_k^A$, we have the following theorem for $\delta_{2k}^A$ and $\delta_{2r}^\mathcal{M}$ .
\begin{Theorem}\label{eq:delta2r}
Suppose $1\leq k\leq p$. Let $y=A\beta$ for a $k$-sparse vector $\beta\in\mathbb{R}^p$. If $\delta_{2k}^A\leq 1/2$, then the minimizer $\hat\beta$ of \eqref{eq:signalmini} with $\mathcal{B}=\{0\}$ recovers $\beta$ exactly, i.e.,  $\hat\beta=\beta$.

Similarly, suppose $1\leq r\leq \min(m,n)$ and let $b=\mathcal{M}(X)$ for some matrix $X$ with $r$-rank. If $\delta_{2r}^{\mathcal{M}}\leq 1/2$, then the minimizer $X_\ast$ of \eqref{eq:matrixmini} with $\mathcal{B}=\{0\}$ recovers $X$ exactly, i.e., $X_\ast=X$.
\end{Theorem}
To the best of our knowledge, these are the best bounds on $\delta_{2k}^A$ and $\delta_{2r}^\mathcal{M}$ available as a sufficient condition for the exact recovery of the sparse signals and low-rank matrices, respectively. Note that Davies and Gribonval \cite{Davies_RIPfail} proved that it is not possible to exactly recover all $k$-sparse signals in the noiseless case when $\delta_{2k}^A> \sqrt{2}/2$. Hence, the upper bounds on $\delta_{2k}^A$ are necessarily less than $\sqrt{2}/2$.
There is still a gap between the two bounds $1/2$ and $\sqrt{2}/2$ on  $\delta_{2k}^A$. It is an interesting future project to close this gap.

It is also noteworthy that Zhang (\cite{Zhang}, Remark 1) proved for some concave penalty $\rho$, the estimator
$$\hat\beta = \arg\min_\beta \left( \|y-X\beta\|_2^2+\sum_{i=1}^p\rho(|\beta_i|, \lambda)\right)$$
 recovers $k$-sparse signals exactly in the noiseless case with a suitable choice of $\lambda$ under the condition $\delta_{2k}<1/2$ or $\delta_{3k}<2/3$. The constrained $\ell_1$ minimization estimator   $\hat\beta$ defined in \eqref{eq:signalmini}  with $\mathcal{B}=\{0\}$ is straightforward to compute. In contrast, the concave penalized minimization estimator requires a good choice of the tuning parameter $\lambda$ and is not as easy to implement.

It is also interesting to consider conditions on $\delta_{sk}^A$ and $\delta_{sr}^\mathcal{M}$ for some integer $s\ge 1$. The following result provides convenient bounds on $\delta_{sk}^A$ and $\delta_{sr}^\mathcal{M}$ in terms of $\delta_{k}^A$ and $\delta_{r}^\mathcal{M}$ respectively. It is also useful for the proof of Theorem \ref{th:oracle}.
\begin{Lemma}\label{lm:delta2k<3deltak}
For all matrix $A\in\mathbb{R}^{n\times p}$ and $k\geq 2$, $s\geq2$, we have $\delta_{sk}^A\leq (2s-1)\delta_k^A$. Similarly, for all linear map $\mathcal{M}:\mathbb{R}^{m\times n}\to\mathbb{R}^q$ and $r\geq 2$, $s\geq 2$, we have $\delta_{sr}^\mathcal{M}\leq (2s-1)\delta_r^\mathcal{M}$.
\end{Lemma}

\section{Proofs}
\label{proofs.sec}

In this section we shall first prove the main results. The proofs of some of the main theorems rely on a few additional technical lemmas. These technical results are collected and proved in Section \ref{lemmas.sec}.

\subsection{Proof of Theorem \ref{th:main}}

The key to the proof of this theorem is parallelogram identity, since it provides equality rather than inequality in the estimation in $\ell_2$ norm as we shall see later.

By $(\ast\ast)$, we only need to show for all $R\in \mathcal{N}(\mathcal{M})\backslash\{0\}$, it satisfies $\|R_{\max(r)}\|_\ast<\|R_{-\max(r)}\|_\ast$.

Suppose there exists $R\in \mathcal{N}(\mathcal{M})\setminus \{0\}$ such that $\|R_{\max(r)}\|_\ast\geq\|R_{-\max(r)}\|_\ast$. Assume $R$ has SVD decomposition $R=\sum_{i=1}^ma_iu_i^Tv_i, a_1\geq a_2\geq \cdots\geq a_m$. Since we can set $a_i=0$ if $i\geq m,n$, without loss of generality we can assume that $m,n \geq r$.

By Lemma \ref{lm:divide}, we can find $\{s_{ij}\}_{1\leq i\leq r,2r+1\leq j\leq m}$ satisfying (\ref{eq:divide1}) and (\ref{eq:divide2}).

\begin{enumerate}

\item When $r$ is even, suppose
\begin{equation}\label{eq:defR11-R23}
\begin{split}
&R_{11}=\sum_{i=1}^{r/2}a_iu_iv_i^T, \quad R_{12}=\sum_{i=r/2+1}^{r}a_iu_iv_i^T,\quad R_{21}=\sum_{i=r+1}^{3r/2}a_iu_iv_i^T, \quad R_{22}=\sum_{i=3r/2+1}^{2r}a_iu_iv_i^T\\
&R_{31}=\sum_{j=2r+1}^m(\sum_{i=1}^{r/2}s_{ij}u_jv_j^T),\quad R_{32}=\sum_{j=2r+1}^m(\sum_{i=r/2+1}^{r}s_{ij}u_jv_j^T)
\end{split}
\end{equation}
then $\mathcal{M}(R_{11}+R_{12}+R_{21}+R_{22}+R_{31}+R_{32})=\mathcal{M}(R)=0$. By the parallelogram identity,
\begin{equation}\label{eq:main}
\begin{split}
&\|\mathcal{M}(-R_{11}+R_{22}+R_{32})\|^2+\|\mathcal{M}(-R_{12}+R_{21}+R_{31})\|^2\\
=&\frac{1}{2}\left[\|\mathcal{M}(-R_{11}-R_{12}+R_{21}+R_{22}+R_{31}+R_{32})\|^2+\|\mathcal{M}(-R_{11}+R_{12}-R_{21}+R_{22}-R_{31}+R_{32})\|^2\right]\\
=&\frac{1}{2}\left[\|\mathcal{M}(2R_{11}+2R_{12})\|^2+\frac{1}{2}\|\mathcal{M}(-2R_{11}-2R_{21}-2R_{31})\|^2+\frac{1}{2}\|\mathcal{M}(2R_{12}+2R_{22}+2R_{32})\|\right]\\
=&2\|\mathcal{M}(R_{11}+R_{12})\|^2+\|\mathcal{M}(R_{11}+R_{21}+R_{31})\|^2+\|\mathcal{M}(R_{12}+R_{22}+R_{32})\|^2
\end{split}
\end{equation}
We use Lemma \ref{lm:AX1-AX2} by setting
$$g=h=r/2,\quad b_i=a_i, c_i=-a_{i+r/2}, d_i=a_{i+r},\quad \forall 1\leq i\leq r/2,$$
$$e_j=\sum_{i=1}^rs_{i,j+2r}, \quad t_{ij}=s_{i,j+2r},\quad 1\leq i\leq r/2, 1\leq j\leq m-2r,$$
then we get
\begin{equation}\label{eq:maininq1}
\begin{split}
&\|\mathcal{M}(R_{11}+R_{21}+R_{31})\|^2-\|\mathcal{M}(-R_{12}+R_{21}+R_{31})\|^2\\
\geq&
(1-\delta_r^{\mathcal{M}})(\sum_{i=1}^{r/2}a_i^2+\sum_{i=r+1}^{3r/2}(a_{i}+\sum_{j=2r+1}^{m}s_{ij})^2)-(1+\delta_r^{\mathcal{M}})(\sum_{i=r/2+1}^{r}a_i^2+\sum_{i=r+1}^{3r/2}(a_{i}+\sum_{j=2r+1}^{m}s_{ij})^2)
\end{split}
\end{equation}
Similarly,
\begin{equation}\label{eq:maininq2}
\begin{split}
&\|\mathcal{M}(R_{12}+R_{22}+R_{32})\|^2-\|\mathcal{M}(-R_{11}+R_{22}+R_{32})\|^2\\
\geq&(1-\delta_r^{\mathcal{M}})(\sum_{i=r/2+1}^{r}a_i^2+\sum_{i=3r/2+1}^{2r}(a_{i}+\sum_{j=2r+1}^{m}s_{ij})^2)-(1+\delta_r^{\mathcal{M}})(\sum_{i=1}^{r/2}a_i^2+\sum_{i=3r/2+1}^{2r}(a_{i}+\sum_{j=2r+1}^{m}s_{ij})^2)
\end{split}
\end{equation}
Let the right hand side of (\ref{eq:main}) minus the left hand side. Along with (\ref{eq:maininq1}), (\ref{eq:maininq2}), we get
\begin{eqnarray*}
0&=&RHS-LHS\\
 &\geq& 2(1-\delta_r^\mathcal{M})(\sum_{i=1}^ra_i^2)-2\delta_r^{\mathcal{M}}\sum_{i=1}^ra_i^2-2\delta_r^{\mathcal{M}}(\sum_{i=r+1}^{2r}(a_{i}+\sum_{j=2r+1}^{m}s_{ij})^2)\\
 &\geq& 2(1-2\delta_r^\mathcal{M})\sum_{i=1}^r a_i^2-2\delta_r^\mathcal{M}r\left(\frac{\sum_{i=1}^ra_i}{r}\right)^2\\
 &\geq& 2(1-3\delta_r^\mathcal{M})\sum_{i=1}^r a_i^2
\end{eqnarray*}
The last two inequalities are due to (\ref{eq:divide2}) and Cauchy-Schwarz inequality. It contradicts the fact that $R\neq 0$ and $\delta_r^{\mathcal{M}}<1/3$.

\item When $r$ is odd, $r\geq 3$, note
\begin{equation}\label{eq:defR11-R33}
\begin{split}
&R_{11}=a_1u_1v_1^T,\quad R_{12}=\sum_{i=2}^{(r+1)/2}a_iu_iv_i^T,\quad R_{13}=\sum_{i=(r+3)/2}^ra_iu_iv_i^T\\
&R_{21}=a_{r+1}u_{r+1}v_{r+1}^T,\quad R_{22}=\sum_{i=r+2}^{(3r+1)/2}a_iu_iv_i^T,\quad R_{23}=\sum_{i=(3r+3)/2}^{2r}a_iu_iv_i^T\\
&R_{31}=\sum_{j=2r+1}^ms_{1j}u_jv_j^T,\quad R_{32}=\sum_{j=2r+1}^m(\sum_{i=2}^{(r+1)/2}s_{ij})u_jv_j^T, \quad R_{33}=\sum_{j=2r+1}^m(\sum_{i=(r+3)/2}^{2r}s_{ij})u_jv_j^T
\end{split}
\end{equation}
Note $X_1=-R_{11}+R_{21}+R_{31}, X_2=-R_{12}+R_{22}+R_{23}, X_3=-R_{13}+R_{23}+R_{33}$, we can easily show the following equality
\begin{equation}\label{eq:4parallelogram}
\begin{split}
&4\|\mathcal{M}(X_1)\|^2+4\|\mathcal{M}(X_2)\|^2+4\|\mathcal{M}(X_3)\|^2\\
=&\|\mathcal{M}(X_1+X_2-X_3)\|^2+\|\mathcal{M}(-X_1+X_2+X_3)\|^2\\
&+\|\mathcal{M}(X_1-X_2+X_3)\|^2+\|\mathcal{M}(X_1+X_2+X_3)\|^2
\end{split}
\end{equation}
By the fact that $\mathcal{M}(R)=0$, (\ref{eq:4parallelogram}) means
\begin{equation}\label{eq:oddequation}
\begin{split}
&\|\mathcal{M}(-R_{11}+R_{21}+R_{31})\|^2+\|\mathcal{M}(-R_{12}+R_{22}+R_{32})\|^2+\|\mathcal{M}(-R_{13}+R_{23}+R_{33})\|^2\\
=&\|\mathcal{M}(R_{12}+R_{13}+R_{21}+R_{31})\|^2+\|\mathcal{M}(R_{11}+R_{13}+R_{22}+R_{32})\|^2\\
&+\|\mathcal{M}(R_{11}+R_{12}+R_{23}+R_{33})\|^2+\|\mathcal{M}(R_{11}+R_{12}+R_{13})\|^2
\end{split}
\end{equation}
Similarly as the even case, by Lemma \ref{lm:AX1-AX2} we have
\begin{equation}\label{eq:mainoddineq1}
\begin{split}
&\|\mathcal{M}(R_{12}+R_{13}+R_{21}+R_{31})\|^2-\|\mathcal{M}(-R_{11}+R_{21}+R_{31})\|^2\\
\geq & (1-\delta_r^{\mathcal{M}})\left[\sum_{i=2}^{r}a_i^2+(a_{r+1}+\sum_{j=2r+1}^ms_{1,j})^2\right]-(1+\delta_r^{\mathcal{M}})\left[a_1^2+(a_{r+1}+\sum_{j=2r+1}^ms_{1,j})^2\right]
\end{split}
\end{equation}
\begin{equation}\label{eq:mainoddineq2}
\begin{split}
&\|\mathcal{M}(R_{11}+R_{13}+R_{22}+R_{32})\|^2-\|\mathcal{M}(-R_{12}+R_{22}+R_{32})\|^2\\
\geq & (1-\delta_r^{\mathcal{M}})\left[a_1^2+\sum_{i=(r+3)/2}^{r}a_i^2+\sum_{i=2}^{(r+1)/2}(a_i+\sum_{j=2r+1}^ms_{ij})^2\right]\\
&-(1+\delta_r^{\mathcal{M}})\left[\sum_{i=2}^{(r+1)/2}a_i^2+\sum_{i=2}^{(r+1)/2}(a_i+\sum_{j=2r+1}^ms_{ij})^2\right]
\end{split}
\end{equation}
\begin{equation}\label{eq:mainoddineq3}
\begin{split}
&\|\mathcal{M}(R_{11}+R_{12}+R_{23}+R_{33})\|^2-\|\mathcal{M}(-R_{13}+R_{23}+R_{33})\|^2\\
\geq & (1-\delta_r^{\mathcal{M}})\left[\sum_{i=1}^{(r+1)/2}a_i^2+\sum_{i=(r+3)/2}^{r}(a_i+\sum_{j=2r+1}^ms_{ij})^2\right]\\
&-(1+\delta_r^{\mathcal{M}})\left[\sum_{i=(r+3)/2}^ra_i^2+\sum_{i=(r+3)/2}^{r}(a_i+\sum_{j=2r+1}^ms_{ij})^2\right]
\end{split}
\end{equation}
Let the right hand side of (\ref{eq:oddequation}) minus the left hand side, we can get
\begin{eqnarray*}
0&\geq&(1-\delta_r^{\mathcal{M}})\left[3\sum_{i=1}^ra_i^2+\sum_{i=1}^r(a_{r+i}+\sum_{j=2r+1}^ms_{ij})^2\right]-(1+\delta_r^{\mathcal{M}})\left[\sum_{i=1}^ra_i^2+\sum_{i=1}^r(a_{r+i}+\sum_{j=2r+1}^ms_{ij})^2\right]\\
 &=&2\left[(1-2\delta_r^{\mathcal{M}})\sum_{i=1}^ra_i^2-\delta_r^{\mathcal{M}}\sum_{i=1}^r(a_{r+i}+\sum_{j=2r+1}^ms_{ij})^2\right]\\
 &\geq&2(1-2\delta_r^{\mathcal{M}})\sum_{i=1}^ra_i^2-2\delta_r^{\mathcal{M}}r\left(\frac{\sum_{i=1}^ra_i}{r}\right)^2\\
 &\geq&2(1-3\delta_r^{\mathcal{M}})\sum_{i=1}^ra_i^2
\end{eqnarray*}
The last two inequalities are due to \eqref{eq:divide2} and Cauchy Schwarz inequality. It contradicts the fact that $R\neq 0$ and $\delta_r^{\mathcal{M}}<1/3$.\quad $\square$

\end{enumerate}

\subsection{Proof of Theorem \ref{th:mainsignal}}

The proof of Theorem \ref{th:mainsignal} is essentially the same as Theorem \ref{th:main}. By $(\ast)$, we only need to show for all $\beta\in \mathcal{N}(A)\setminus\{0\}$, it satisfies $\|\beta_{\max(k)}\|_1<\|\beta_{-\max(k)}\|_1$.

For the convenience of presentation, we call a vector with 1 or -1 in only one entry and zeros elsewhere as the \emph{indicator vector}.

Suppose there exists $\beta\in \mathcal{N}(\mathcal{A})\setminus \{0\}$ such that $\|\beta_{\max(k)}\|_1<\|\beta_{-\max(k)}\|_1$. Then $\beta$ can be written as
$$\beta=\sum_{i=1}^p a_iu_i$$
where $\{u_i\}_{i=1}^p$ are indicator vectors with different support in $\mathbb{R}^p$; $\{a_i\}_{i=1}^p$ is a non-negative and decreasing sequence. Since we can set $a_i=0$ if $i\geq p$, without loss of generality we can assume that $p \geq k$.

By Lemma \ref{lm:divide}, we can find $\{s_{ij}\}_{1\leq i\leq k,2k+1\leq j\leq p}$ satisfying (\ref{eq:divide1}) and (\ref{eq:divide2}) with a modification of notations.

\begin{enumerate}

\item When $k$ is even, suppose
\begin{equation}\label{eq:defR11-R23signal}
\begin{split}
&\beta_{11}=\sum_{i=1}^{k/2}a_iu_i, \quad \beta_{12}=\sum_{i=k/2+1}^{k}a_iu_i,\quad \beta_{21}=\sum_{i=k+1}^{3k/2}a_iu_i, \quad \beta_{22}=\sum_{i=3k/2+1}^{2k}a_iu_i\\
&\beta_{31}=\sum_{j=2k+1}^p(\sum_{i=1}^{k/2}s_{ij}u_j),\quad \beta_{32}=\sum_{j=2k+1}^p(\sum_{i=k/2+1}^{k}s_{ij}u_j)
\end{split}
\end{equation}
then $A(\beta_{11}+\beta_{12}+\beta_{21}+\beta_{22}+\beta_{31}+\beta_{32})=A\beta=0$. By the parallelogram identity,
\begin{equation}\label{eq:mainsignal}
\begin{split}
&\|A(-\beta_{11}+\beta_{22}+\beta_{32})\|^2+\|A(-\beta_{12}+\beta_{21}+\beta_{31})\|^2\\
=&\frac{1}{2}\left[\|A(-\beta_{11}-\beta_{12}+\beta_{21}+\beta_{22}+\beta_{31}+\beta_{32})\|^2+\|A(-\beta_{11}+\beta_{12}-\beta_{21}+\beta_{22}-\beta_{31}+\beta_{32})\|^2\right]\\
=&\frac{1}{2}\left[\|A(2\beta_{11}+2\beta_{12})\|^2+\frac{1}{2}\|A(-2\beta_{11}-2\beta_{21}-2\beta_{31})\|^2+\frac{1}{2}\|A(2\beta_{12}+2\beta_{22}+2\beta_{32})\|\right]\\
=&2\|A(\beta_{11}+\beta_{12})\|^2+\|A(\beta_{11}+\beta_{21}+\beta_{31})\|^2+\|A(\beta_{12}+\beta_{22}+\beta_{32})\|^2
\end{split}
\end{equation}
Similarly as the matrix case, we use Lemma \ref{lm:AX1-AX2} and get
\begin{equation}\label{eq:maininq1signal}
\begin{split}
&\|A(\beta_{11}+\beta_{21}+\beta_{31})\|^2-\|A(-\beta_{12}+\beta_{21}+\beta_{31})\|^2\\
\geq&
(1-\delta_k^{A})(\sum_{i=1}^{k/2}a_i^2+\sum_{i=k+1}^{3k/2}(a_{i}+\sum_{j=2k+1}^{p}s_{ij})^2)-(1+\delta_k^A)(\sum_{i=k/2+1}^{k}a_i^2+\sum_{i=k+1}^{3k/2}(a_{i}+\sum_{j=2k+1}^{p}s_{ij})^2)
\end{split}
\end{equation}
Similarly,
\begin{equation}\label{eq:maininq2signal}
\begin{split}
&\|A(\beta_{12}+\beta_{22}+\beta_{32})\|^2-\|A(-\beta_{11}+\beta_{22}+\beta_{32})\|^2\\
\geq&(1-\delta_k^{A})(\sum_{i=k/2+1}^{k}a_i^2+\sum_{i=3k/2+1}^{2k}(a_{i}+\sum_{j=2k+1}^{p}s_{ij})^2)-(1+\delta_k^{A})(\sum_{i=1}^{k/2}a_i^2+\sum_{i=3k/2+1}^{2k}(a_{i}+\sum_{j=2k+1}^{p}s_{ij})^2)
\end{split}
\end{equation}
Let the right hand side of (\ref{eq:mainsignal}) minus the left hand side. Along with (\ref{eq:maininq1signal}), (\ref{eq:maininq2signal}), we get
\begin{eqnarray*}
0&=&RHS-LHS\\
 &\geq& 2(1-\delta_k^A)(\sum_{i=1}^ka_i^2)-2\delta_k^A\sum_{i=1}^ka_i^2-2\delta_k^A(\sum_{i=k+1}^{2k}(a_{i}+\sum_{j=2k+1}^{p}s_{ij})^2)\\
 &\geq& 2(1-2\delta_k^A)\sum_{i=1}^k a_i^2-2\delta_k^A k\left(\frac{\sum_{i=1}^ka_i}{k}\right)^2\\
 &\geq& 2(1-3\delta_k^A)\sum_{i=1}^k a_i^2
\end{eqnarray*}
The last two inequalities are due to (\ref{eq:divide2}) and Cauchy-Schwarz inequality. It contradicts the fact that $\beta\neq 0$ and $\delta_k^A<1/3$.

\item When $k$ is odd, $k\geq 3$, note
\begin{equation}\label{eq:defR11-R33signal}
\begin{split}
&\beta_{11}=a_1u_1,\quad \beta_{12}=\sum_{i=2}^{(k+1)/2}a_iu_i,\quad \beta_{13}=\sum_{i=(k+3)/2}^ka_iu_i\\
&\beta_{21}=a_{k+1}u_{k+1},\quad \beta_{22}=\sum_{i=k+2}^{(3k+1)/2}a_iu_i,\quad \beta_{23}=\sum_{i=(3k+3)/2}^{2k}a_iu_i\\
&\beta_{31}=\sum_{j=2k+1}^ps_{1j}u_j,\quad \beta_{32}=\sum_{j=2k+1}^p(\sum_{i=2}^{(k+1)/2}s_{ij})u_j, \quad \beta_{33}=\sum_{j=2k+1}^p(\sum_{i=(k+3)/2}^{2k}s_{ij})u_j
\end{split}
\end{equation}
Note $\gamma_1=-\beta_{11}+\beta_{21}+\beta_{31}, \gamma_2=-\beta_{12}+\beta_{22}+\beta_{23}, \gamma_3=-\beta_{13}+\beta_{23}+\beta_{33}$, we can easily show the following equality
\begin{equation}\label{eq:4parallelogramsignal}
\begin{split}
&4\|A\gamma_1\|^2+4\|A\gamma_2\|^2+4\|A\gamma_3\|^2\\
=&\|A(\gamma_1+\gamma_2-\gamma_3)\|^2+\|A(-\gamma_1+\gamma_2+\gamma_3)\|^2\\
&+\|A(\gamma_1-\gamma_2+\gamma_3)\|^2+\|A(\gamma_1+\gamma_2+\gamma_3)\|^2
\end{split}
\end{equation}
By the fact that $A\beta=0$, (\ref{eq:4parallelogramsignal}) means
\begin{equation}\label{eq:oddequationsignal}
\begin{split}
&\|A(-\beta_{11}+\beta_{21}+\beta_{31})\|^2+\|A(-\beta_{12}+\beta_{22}+\beta_{32})\|^2+\|A(-\beta_{13}+\beta_{23}+\beta_{33})\|^2\\
=&\|A(\beta_{12}+\beta_{13}+\beta_{21}+\beta_{31})\|^2+\|A(\beta_{11}+\beta_{13}+\beta_{22}+\beta_{32})\|^2\\
&+\|A(\beta_{11}+\beta_{12}+\beta_{23}+\beta_{33})\|^2+\|A(\beta_{11}+\beta_{12}+\beta_{13})\|^2
\end{split}
\end{equation}
Similarly as the even case, by Lemma \ref{lm:AX1-AX2} we have
\begin{equation}\label{eq:mainoddineq1signal}
\begin{split}
&\|A(\beta_{12}+\beta_{13}+\beta_{21}+\beta_{31})\|^2-\|A(-\beta_{11}+\beta_{21}+\beta_{31})\|^2\\
\geq & (1-\delta_k^A)\left[\sum_{i=2}^{k}a_i^2+(a_{k+1}+\sum_{j=2k+1}^ps_{1,j})^2\right]-(1+\delta_k^A)\left[a_1^2+(a_{k+1}+\sum_{j=2k+1}^ps_{1,j})^2\right]
\end{split}
\end{equation}
\begin{equation}\label{eq:mainoddineq2signal}
\begin{split}
&\|A(\beta_{11}+\beta_{13}+\beta_{22}+\beta_{32})\|^2-\|A(-\beta_{12}+\beta_{22}+\beta_{32})\|^2\\
\geq & (1-\delta_k^A)\left[a_1^2+\sum_{i=(k+3)/2}^{k}a_i^2+\sum_{i=2}^{(k+1)/2}(a_i+\sum_{j=2k+1}^ps_{ij})^2\right]\\
&-(1+\delta_k^A)\left[\sum_{i=2}^{(k+1)/2}a_i^2+\sum_{i=2}^{(k+1)/2}(a_i+\sum_{j=2k+1}^ps_{ij})^2\right]
\end{split}
\end{equation}
\begin{equation}\label{eq:mainoddineq3signal}
\begin{split}
&\|A(\beta_{11}+\beta_{12}+\beta_{23}+\beta_{33})\|^2-\|A(-\beta_{13}+\beta_{23}+\beta_{33})\|^2\\
\geq & (1-\delta_k^A)\left[\sum_{i=1}^{(k+1)/2}a_i^2+\sum_{i=(k+3)/2}^{k}(a_i+\sum_{j=2k+1}^ps_{ij})^2\right]\\
&-(1+\delta_k^A)\left[\sum_{i=(k+3)/2}^ka_i^2+\sum_{i=(k+3)/2}^{k}(a_i+\sum_{j=2k+1}^ps_{ij})^2\right]
\end{split}
\end{equation}
Let the right hand side of (\ref{eq:oddequationsignal}) minus the left hand side, we can get
\begin{eqnarray*}
0&\geq&(1-\delta_k^A)\left[3\sum_{i=1}^ka_i^2+\sum_{i=1}^k(a_{k+i}+\sum_{j=2k+1}^ps_{ij})^2\right]-(1+\delta_k^A)\left[\sum_{i=1}^ka_i^2+\sum_{i=1}^k(a_{k+i}+\sum_{j=2k+1}^ps_{ij})^2\right]\\
 &=&2\left[(1-2\delta_k^A)\sum_{i=1}^ka_i^2-\delta_k^A\sum_{i=1}^k(a_{k+i}+\sum_{j=2k+1}^ps_{ij})^2\right]\\
 &\geq&2(1-2\delta_k^A)\sum_{i=1}^ka_i^2-2\delta_k^Ak\left(\frac{\sum_{i=1}^ka_i}{k}\right)^2\\
 &\geq&2(1-3\delta_k^A)\sum_{i=1}^ka_i^2
\end{eqnarray*}
The last two inequalities are due to \eqref{eq:divide2} and Cauchy Schwarz inequality. It contradicts the fact that $\beta\neq 0$ and $\delta_k^A<1/3$.\quad $\square$

\end{enumerate}

\subsection{Proof of Theorem \ref{th:counterexample}}

It is well known that for matrices $X$, $B$ with the same size, $|\langle X,B \rangle|\leq\|X\|_F\|B\|_F$. The following lemma provides a stronger result given further constraint on matrix rank.
\begin{Lemma}\label{lm:counterexample}
Let $X\in \mathbb{R}^{m\times n} (m\leq n)$ be a matrix with singular values $\lambda_1\geq\lambda_2\geq\cdots\geq\lambda_m$, then for all $B\in \mathbb{R}^{m\times n}$ such that $rank(B)\leq r$, we have
$$|\langle B,X\rangle|\leq\|B\|_F\sqrt{\sum_{i=1}^r\lambda_i^2}.$$
\end{Lemma}

\noindent\textbf{Proof of Lemma \ref{lm:counterexample}}  Since the rank of $B$ is at most $r$, we can suppose $B,X$ have singular value decomposition $B=U\Sigma V$,$X=W\Lambda Z$, where $U,W\in \mathbb{R}^{m\times m}, \Sigma,\Lambda\in \mathbb{R}^{m\times n}, V,Z\in \mathbb{R}^{n\times n}$.
Then
$$\langle B,X\rangle=tr(B^TX)=tr(V^T\Sigma^TU^TW\Lambda Z)=tr(\Sigma^TU^TW\Lambda ZV^T)=\diag(\Sigma)\cdot diag(U^TW\Lambda ZV^T)$$
Since the rank of $B$ is at most $r$, $diag(\Sigma)$ is supported on the first $r$ entries,
$$|\langle B,X\rangle|\leq \sqrt{\sum_{i=1}^r\Sigma_{ii}^2}\cdot\sqrt{\sum_{i=1}^r(U^TW\Lambda ZV^T)^2_{ii}}\leq\| B\|_F\sqrt{\sum_{i=1}^r\sum_{j=1}^n(U^TW\Lambda ZV^T)^2_{ij}}=\|B\|_F\|K\Lambda ZV^T\|_F$$
where we note $K\in \mathbb{R}^{r\times n}$ as the first $r$ rows of $U^TW$. In addition,
$$\|K\Lambda ZV^T\|_F^2=tr(VZ^T\Lambda^T K^TK\Lambda ZV^T)=tr(\Lambda ZV^TVZ^T\Lambda^T K^TK)=tr(\Lambda^2K^TK)$$
By $K$ is the first $r$ row of an $n\times n$ orthogonal matrix, we have $tr(K^TK)=tr(KK^T)=tr(I_r)=r$ and all diagonal elements of $K^TK$ are in $[0,1]$, then
$$tr(\Lambda^2K^TK)=\sum_{i=1}^n\lambda_i^2(K^TK)_{ii}\leq \sum_{i=1}^r\lambda_i^2$$
In summary,
$$|\langle B,X\rangle|\leq \|B\|_F\|K\Lambda ZV^T\|_F\leq\|B\|_F\sqrt{\sum_{i=1}^r\lambda_i^2}.\quad\square$$

It is noteworthy that the signal version of this lemma simply holds by Cauchy-Schwarz inequality.

Now we construct an example for Theorem \ref{th:counterexample}, then check the feasibility by the lemma above. Note
$$X_1=\diag(\overbrace{\frac{1}{\sqrt{2r}},\cdots,\frac{1}{\sqrt{2r}}}^{2r},0,\cdots,0)\in \mathbb{R}^{m\times n}$$
Suppose $H=(\mathbb{R}^{m\times n},\|X\|_F)$ is the Hilbert with inner product $\langle\cdot,\cdot\rangle$. Since $\|X_1\|_F=1$, we can extend $X_1$ into a basis $\{X_1,\cdots,X_{mn} \}$. Define $\mathcal{M}:\mathbb{R}^{m\times n}\to \mathbb{R}^{mn}$ as
\begin{equation}\label{eq:counterexampleA}
\mathcal{M}(X)=\sqrt{\frac{4}{3}}\sum_{i=2}^{mn}a_iX_i
\end{equation}
for all $X=\sum_{i=1}^{mn}a_iX_i$.

Then by Lemma \ref{lm:counterexample}, for all matrix $X$ with rank at most $r$, we have
$$|\langle X,X_1\rangle|\leq\sqrt{r\cdot\frac{1}{2r}}\|X\|_F= \sqrt{\frac{1}{2}}\|X\|_F$$
$$\|\mathcal{M}(X)\|_2^2=\frac{4}{3}\sum_{i=2}^{mn}a_i^2=\frac{4}{3}(\|X\|_F^2-a_1^2)=\frac{4}{3}(\|X\|_F^2-|\langle X,X_1\rangle|^2)$$
Thus,
$$\frac{2}{3}\|X\|_F^2\leq\|\mathcal{M}(X)\|_2^2\leq\frac{4}{3}\|X\|_F^2,\quad \delta_r^\mathcal{M}(X)\leq 1/3$$
Notice that
$$\|\mathcal{M}(\diag(\overbrace{1,\cdots,1}^r,0,\cdots,0))\|^2_2=\frac{2}{3}r=\frac{2}{3}\|(\diag(\overbrace{1,\cdots,1}^r,0,\cdots,0))\|_F^2$$
$$\|\mathcal{M}(\diag(1,-1,0,\cdots,0))\|^2_2=\frac{8}{3}=\frac{4}{3}\|\diag(1,-1,0,\cdots,0)\|_F^2$$
we can conclude that $\delta_r^\mathcal{M}=1/3$. Finally, suppose
$$X=\diag(\overbrace{1,1\cdots,1}^r,0,\cdots,0),\quad Y=\diag(\overbrace{0,\cdots,0}^r, \overbrace{-1,-1\cdots,-1}^r,0,\cdots,0)$$
Then $X$, $Y$ are both matrices of rank $r$ such that $X-Y\in\mathcal{N}(\mathcal{M})$, $\mathcal{M}(X)=\mathcal{M}(Y)$.
Therefore, it is impossible to recover both $X$ and $Y$ only given $(b,\mathcal{M})$, which finishes the proof. $\quad \square$

\subsection{Proof of Theorems \ref{th:counterexamplesignal}}

Again, the proof to this theorem is essentially the same as Theorem \ref{th:counterexample}. Note
$$\beta_1=(\overbrace{\frac{1}{\sqrt{2k}},\cdots,\frac{1}{\sqrt{2k}}}^{2k},0,\cdots,0)\in \mathbb{R}^{p}$$
Suppose $H=(\mathbb{R}^{p},\|\cdot\|_2)$ is the Hilbert with inner product $\langle\cdot,\cdot\rangle$. Since $\|\beta_1\|_2=1$, we can extend $\beta_1$ into a basis $\{\beta_1,\cdots,\beta_p \}$. Define $A:\mathbb{R}^{p}\to \mathbb{R}^{p}$ as
\begin{equation}\label{eq:counterexampleAsignal}
A\beta=\sqrt{\frac{4}{3}}\sum_{i=2}^{p}a_i\beta_i
\end{equation}
for all $\beta=\sum_{i=1}^{p}a_i\beta_i$.

Then by Cauchy-Schwarz inequality, for all $k$-sparse vector $\gamma$, we have
$$|\langle \gamma,\beta_1\rangle|\leq\|\beta_1\cdot1_{\textrm{supp}(\gamma)}\|_2\|\gamma\|_2\leq \sqrt{\frac{1}{2}}\|\gamma\|_2$$
$$\|A\gamma\|_2^2=\frac{4}{3}\sum_{i=2}^{p}a_i^2=\frac{4}{3}(\|\gamma\|_2^2-a_1^2)=\frac{4}{3}(\|\gamma\|_2^2-|\langle \gamma,\beta_1\rangle|^2)$$
Thus,
$$\frac{2}{3}\|\gamma\|_2^2\leq\|A\gamma\|_2^2\leq\frac{4}{3}\|\gamma\|_2^2,\quad \delta_k^A\leq 1/3$$
Notice that
$$\|A(\overbrace{1,\cdots,1}^k,0,\cdots,0)\|^2_2=\frac{2}{3}k=\frac{2}{3}\|(\overbrace{1,\cdots,1}^k,0,\cdots,0)\|_2^2$$
$$\|A(1,-1,0,\cdots,0)\|^2_2=\frac{8}{3}=\frac{4}{3}\|(1,-1,0,\cdots,0)\|_2^2$$
we can conclude that $\delta_k^A=1/3$. Finally, suppose
$$\gamma=(\overbrace{1,1\cdots,1}^k,0,\cdots,0),\quad \eta=(\overbrace{0,\cdots,0}^k, \overbrace{-1,-1\cdots,-1}^k,0,\cdots,0)$$
Then $\gamma$, $\eta$ are both matrices of rank $k$ such that $\gamma-\eta\in\mathcal{N}(A)$, $A\gamma=A\eta$.
Therefore, it is impossible to recover both $\gamma$ and $\eta$ only given $(y,A)$, which finishes the proof. $\quad \square$

\subsection{Proof of Theorems \ref{th:noisysignal} and \ref{th:noisy}}

For the proof of Theorem \ref{th:noisysignal} and Theorem \ref{th:noisy}, we only show the the latter one about the matrix case, as the proof to the signal case is similar and simpler. Suppose $R=X_\ast-X$, $h=\hat\beta-\beta$. We will use a widely used fact. The readers may see \cite{Cai_l1}, \cite{Candes_Dantzig}, \cite{Candes_incompletemeasurements}, \cite{Donoho} (signal case) or \cite{Wang_NewRIC} (matrix case) for details:
$$\|h_{-\max(k)}\|_1\leq \|h_{\max(k)}\|_\ast+2\|\beta_{-\max(k)}\|_1$$
$$\|R_{-\max(r)}\|_\ast\leq \|R_{\max(r)}\|+2\|X_{-\max(r)}\|_\ast$$
For the remaining part of proof, we only prove the matrix case. Suppose $R$ has singular value decomposition $R=\sum_{i=1}^ma_iu_iv_i^T$. Then we have
\begin{equation}\label{eq:a1+X_0c>=am}
\sum_{i=1}^ra_i+2\|X_{-max(r)}\|_\ast\geq \sum_{i=r+1}^ma_i
\end{equation}
Apply Division Lemma \ref{lm:divide} by setting $a'_i=a_i+2\|X_{-max(r)}\|_\ast/r, i=1,\cdots,r$ and $a_j'=a_j, j>r+1$, we can find $\{s_{ij}\}_{1\leq i\leq r,2r+1\leq j\leq m}$ satisfying
\begin{equation}\label{eq:dividegeneral1}
\sum_{i=1}^rs_{ij}=a_j,\quad \forall \; 2r+1\leq j\leq m,
\end{equation}
\begin{equation}\label{eq:dividegeneral2}
\frac{1}{r}\sum_{w=1}^ra_{w}+\frac{2\|X_{-max(r)}\|_\ast}{r}\geq a_{r+i}+\sum_{j=2r+1}^m s_{ij} ,\quad \forall \; 1\leq i\leq r.
\end{equation}
We also know
\begin{equation}\label{eq:ineqA(R)}
\|\mathcal{M}(R)\|\leq\|\mathcal{M}(X)-b\|+\|b-\mathcal{M}(X_\ast)\|\leq\varepsilon+\eta.
\end{equation}
Similarly as Theorem \ref{th:main}, we finish the remaining part of proof for even or odd $r$ separately.
\begin{enumerate}
\item When $r$ is even, we define $R_{11},\cdots,R_{32}$ as (\ref{eq:defR11-R23}), similarly as (\ref{eq:main}) and by parallelogram equality, we get
\begin{equation}\label{eq:noisyevenmain}
\begin{split}
&\|\mathcal{M}(-R_{11}+R_{22}+R_{32})\|^2+\|\mathcal{M}(-R_{12}+R_{21}+R_{31})\|^2\\
=&\frac{1}{2}\big[\|\mathcal{M}(-R_{11}-R_{12}+R_{21}+R_{22}+R_{31}+R_{32})\|^2\\
&+\|\mathcal{M}(-R_{11}+R_{12}-R_{21}+R_{22}-R_{31}+R_{32})\|^2\big]\\
=&\frac{1}{2}\|\mathcal{M}(2R_{11}+2R_{12})-\mathcal{M}(R)\|^2+\frac{1}{4}\|\mathcal{M}(-2R_{11}-2R_{21}-2R_{31})\|^2\\
&+\frac{1}{4}\|\mathcal{M}(2R_{12}+2R_{22}+2R_{32})\|^2-\frac{1}{8}\|\mathcal{M}(2R)\|^2\\
=&2\|\mathcal{M}(R_{11}+R_{12})\|^2+\|\mathcal{M}(R_{11}+R_{21}+R_{31})\|^2\\
&+\|\mathcal{M}(R_{12}+R_{22}+R_{32})\|^2-2\langle\mathcal{M}(R),\mathcal{M}(R_{11}+R_{12})\rangle
\end{split}
\end{equation}
Let the right hand side of (\ref{eq:noisyevenmain}) minus the left hand side. Along with (\ref{eq:maininq1}), (\ref{eq:maininq2}), one get
\begin{equation}\label{eq:ineqlast}
\begin{split}
0&=RHS-LHS\\
 &\geq 2(1-\delta_r^\mathcal{M})\sum_{i=1}^ra_i^2-2\delta_r^{\mathcal{M}}\sum_{i=1}^ra_i^2-2\delta_r^{\mathcal{M}}(\sum_{i=r+1}^{2r}(a_{i}+\sum_{j=2r+1}^{m}s_{ij})^2)-2\langle\mathcal{M}(R),\mathcal{M}(R_{11}+R_{12})\rangle\\
 &\geq 2(1-2\delta_r^\mathcal{M})\sum_{i=1}^ra_i^2-2\delta_r^\mathcal{M}r(\frac{\sum_{i=1}^ra_{i}}{r}+\frac{2\|X_{-max(r)}\|_\ast}{r})^2-2(\varepsilon+\eta)\sqrt{(1+\delta_r^\mathcal{M})\sum_{i=1}^r a_i^2}\\
 &\geq 2(1-2\delta_r^\mathcal{M})\sum_{i=1}^ra_i^2-2\delta_r^\mathcal{M}(\sqrt{\sum_{i=1}^ra_i^2}+\frac{2\|X_{-max(r)}\|_\ast}{\sqrt{r}})^2-2(\varepsilon+\eta)\sqrt{(1+\delta_r^\mathcal{M})\sum_{i=1}^r a_i^2}
\end{split}
\end{equation}
By \eqref{eq:ineqlast} we can get an inequality of $\sqrt{\sum_{i=1}^ra_i^2}$:
\begin{equation}
\begin{split}
\sqrt{\sum_{i=1}^ra_i^2}\leq& \frac{\delta\frac{2\|X_{-max(r)}\|_\ast}{\sqrt{r}}+\frac{\varepsilon+\eta}{2}\sqrt{1+\delta}}{1-3\delta}\\
&+\frac{\sqrt{(\delta\frac{2\|X_{-max(r)}\|_\ast}{\sqrt{r}}+\frac{\varepsilon+\eta}{2}\sqrt{1+\delta})^2+(1-3\delta)\delta\|2X_{-max(r)}\|_\ast^2/r}}{1-3\delta}\\
\leq& \frac{\sqrt{1+\delta}(\varepsilon+\eta)+2(2\delta+\sqrt{(1-3\delta)\delta})\|X_{-max(r)}\|_\ast/\sqrt{r}}{1-3\delta}
\end{split}
\end{equation}
Finally, by Lemma \ref{lm:ineq},
$$\sum_{i=r+1}^ma_i^2\leq
(\sqrt{\sum_{i=1}^ra_i^2}+\frac{2\|X_{-max(r)}\|_\ast}{\sqrt{r}})^2$$
Then
\begin{equation}
\begin{split}
\|R\|_F&=\sqrt{\sum_{i=1}^m a_i^2}\leq \sqrt{\sum_{i=1}^ra_i^2+(\sqrt{\sum_{i=1}^ra_i^2}+\frac{2\|X_{-max(r)}\|_\ast}{\sqrt{r}})^2}\leq\sqrt{2\sum_{i=1}^ra_i^2}+\frac{2\|X_{-max(r)}\|}{\sqrt{r}}\\
 &\leq \frac{\sqrt{2(1+\delta)}}{1-3\delta}(\varepsilon+\eta)+\frac{2\sqrt{2}(2\delta+\sqrt{(1-3\delta)\delta})+2(1-3\delta)}{1-3\delta}\frac{\|X_{-max(r)}\|_\ast}{\sqrt{r}}
\end{split}
\end{equation}

\item When $r$ is odd, we use the definitions in \eqref{eq:defR11-R33}. Similar equality as \eqref{eq:noisyevenmain} holds as follows,
\begin{equation*}
\begin{split}
&\|\mathcal{M}(-R_{11}+R_{21}+R_{31})\|^2+\|\mathcal{M}(-R_{12}+R_{22}+R_{32})\|^2+\|\mathcal{M}(-R_{13}+R_{23}+R_{33})\|^2\\
=&\|\mathcal{M}(R_{12}+R_{13}+R_{21}+R_{31})\|^2+\|\mathcal{M}(R_{11}+R_{13}+R_{22}+R_{32})\|^2\\
&+\|\mathcal{M}(R_{11}+R_{12}+R_{23}+R_{33})\|^2+\|\mathcal{M}(R_{11}+R_{12}+R_{13})\|^2\\
&-2\langle\mathcal{M}(R_{11}+R_{12}+R_{13},\mathcal{M}(R)\rangle
\end{split}
\end{equation*}
By the method in the even case, we can still get the inequality (\ref{eq:ineqlast}). Hence we have the same estimation.\quad $\square$
\end{enumerate}

\subsection{Proof of Theorem \ref{eq:delta2r}}

By ($\ast$), we only need to show for all $R\in \mathcal{N}(\mathcal{M})\backslash\{0\}$, it satisfies $\|R_{\max(r)}\|_\ast<\|R_{-max(r)}\|_\ast$.

 Suppose there exists $R\in \mathcal{N}(\mathcal{M})\backslash\{0\}$ such that $\|R_{\max(r)}\|_\ast\geq\|R_{-\max(r)}\|_\ast$. Suppose $R$ has singular value decomposition: $\sum_{i=1}^m a_iu_iv_i^T$. Note:
\begin{equation}
R_1=\sum_{i=1}^r a_iu_iv_i^T, \quad R_2=\sum_{i=r+1}^{2r} a_iu_iv_i^T,\quad R_3=\sum_{i=2r+1}^{3r}a_iu_iv_i^T,\quad R_c=\sum_{i=3r+1}^ma_iu_iv_i^T\\
\end{equation}
Notice that $\sum_{i=1}^ra_i\geq\sum_{i=r+1}^{m}a_i\geq\sum_{i=2r+1}^ma_i$. In addition, two equalities cannot hold simultaneously since $R\neq0$. Thus,
$$\sum_{i=1}^ra_i>\sum_{i=2r+1}^ma_i.$$
 Apply Lemma \ref{lm:divide} to $\{a_1,\cdots,a_r,a_{2r+1},\cdots,a_m\}$, we can find $\{s_{ij}\}_{1\leq i\leq r, 3r+1\leq j\leq m}$ such that
$$\sum_{i=1}^rs_{ij}=a_{j},\quad \forall 3r+1\leq j\leq m; \quad\frac{\sum_{w=1}^ra_{w}}{r}\geq a_{2r+i}+ \sum_{j=3r+1}^ms_{ij},\quad \forall 1\leq i\leq r$$
By $\sum_{i=1}^ra_i>\sum_{i=2r+1}^ma_i$, there exists $1\leq i\leq r$ such that $\frac{\sum_{w=1}^ra_{w}}{r}> a_{2r+i}+ \sum_{j=3r+1}^ms_{ij}$. We also have the equality in $l_2$ space as follows,
\begin{equation}\label{eq:delta2requality}
\begin{split}
6\|\mathcal{M}(R_1+R_2)\|^2+3\|\mathcal{M}(R_1+R_3+R_c)\|^2&=2\|\mathcal{M}(-R_2+R_3+R_c)\|^2+\|\mathcal{M}(3R_1+2R_2+R_3+R_c)\|^2\\
 &=2\|\mathcal{M}(-R_2+R_3+R_c)\|^2+\|\mathcal{M}(-R_1+R_3+R_c)\|^2
\end{split}
\end{equation}
Let the left hand side of \eqref{eq:delta2requality} minus the right hand side, by Lemma \ref{lm:AX1-AX2} we get
\begin{eqnarray*}
0 &  = & 6\|\mathcal{M}(R_1+R_2)\|^2+ 2\left(\|\mathcal{M}(R_1+R_3+R_c)\|^2-\|\mathcal{M}(-R_2+R_3+R_c)\|^2\right)\\
  &    &+\left(\|\mathcal{M}(R_1+R_3+R_c)\|^2-\|\mathcal{M}(-R_1+R_3+R_c)\|^2\right)\\
  & \geq & 6(1-\delta_{2r}^\mathcal{M})\sum_{i=1}^{2r}a_i^2+(1-\delta_{2r}^\mathcal{M})\left(2\sum_{i=1}^ra_i^2+3\sum_{i=1}^r(a_{2r+i}+\sum_{j=3r+1}^ms_{ij})^2+\sum_{i=1}^ra_i^2\right)\\
  & &-(1+\delta_{2r}^\mathcal{M})\left(2\sum_{i=r+1}^{2r}a_i^2+3\sum_{i=1}^r(a_{2r+i}+\sum_{j=3r+1}^ms_{ij})^2+\sum_{i=1}^ra_i^2\right)\\
  & = & (8-10\delta_{2r}^\mathcal{M})\sum_{i=1}^ra_i^2+(4-8\delta_{2r}^\mathcal{M})\sum_{r+1}^{2r}a_i^2-6\delta_{2r}^\mathcal{M}\sum_{i=1}^r(a_{2r+i}+\sum_{j=3r+1}^ms_{ij})^2\\
  & \geq & 3\left(\sum_{i=1}^ra_i^2-\sum_{i=1}^r(a_{2r+i}+\sum_{j=3r+1}^ms_{ij})^2\right)\\
  & > & 3\left(\sum_{i=1}^ra_i^2-r(\frac{\sum_{i=1}^ra_i}{r})^2\right)\geq0
\end{eqnarray*}
which is a contradiction.\quad $\square$

\subsection{Proof of Lemma \ref{lm:delta2k<3deltak}}
We only show the matrix case. For all $X\in\mathbb{R}^{m\times n}$ such that $rank(X)\leq 2r$, suppose $X$ has singular value decomposition $X=\sum_{i=1}^la_iu_iv_i^T$, $l\leq sr$. Without loss of generality we can assume $l=sr$ as we can set $a_i=0$ if $l<i\leq sr$. Note
$$w_i=\mathcal{M}(a_iu_iv_i^T)\in\mathbb{R}^q, \quad 1\leq i \leq sr$$
We can verify the following identity
\begin{eqnarray*}
& &\|\sum_{i=1}^{sr}w_i\|_2^2+\frac{s-1}{sr-1}\sum_{1\leq i<j\leq sr}\|w_i-w_j\|_2^2\\
&=&(1+(s-1))\sum_{i=1}^{sr}\|w_i\|_2^2+2(1-\frac{s-1}{sr-1})\sum_{1\leq i<j\leq sr}\langle w_i,w_j\rangle\\
&=&\frac{s^2}{\binom{sr}{r}}\sum_{1\leq i_1<\cdots<i_r\leq sr}\|w_{i_1}+w_{i_2}+\cdots+w_{i_r}\|_2^2
\end{eqnarray*}
which implies
\begin{eqnarray*}
\|\mathcal{M}(X)\|_2^2&=&\|\sum_{i=1}^{sr}w_i\|_2^2\\
&\leq& \frac{s^2(1+\delta_r^\mathcal{M})}{\binom{sr}{r}}\sum_{1\leq i_1<\cdots<i_r\leq sr}(a_{i_1}^2+\cdots+a_{i_r}^2)-\frac{(s-1)(1-\delta_r^\mathcal{M})}{sr-1}\sum_{1\leq i<j\leq sr}(a_i^2+a_j^2)\\
&=& (s(1+\delta_r^\mathcal{M})-(s-1)(1-\delta_r^\mathcal{M}))\sum_{i=1}^{rs}a_i^2\\
&=& (1+(2s-1)\delta_r^\mathcal{M})\|X\|_F^2
\end{eqnarray*}
\begin{eqnarray*}
\|\mathcal{M}(X)\|_2^2&=&\|\sum_{i=1}^{sr}w_i\|_2^2\\
&\geq& \frac{s^2(1-\delta_r^\mathcal{M})}{\binom{sr}{r}}\sum_{1\leq i_1<\cdots<i_r\leq sr}(a_{i_1}^2+\cdots+a_{i_r}^2)-\frac{(s-1)(1+\delta_r^\mathcal{M})}{sr-1}\sum_{1\leq i<j\leq sr}(a_i^2+a_j^2)\\
&=& (s(1-\delta_r^\mathcal{M})-(s-1)(1+\delta_r^\mathcal{M}))\sum_{i=1}^{rs}a_i^2\\
&=& (1-(2s-1)\delta_r^\mathcal{M})\|X\|_F^2
\end{eqnarray*}
Hence, $\delta_{sr}^\mathcal{M}\leq (2s-1)\delta_{r}^\mathcal{M}$. \quad $\square$

\subsection{Proof of Theorem \ref{th:oracle}}

By a small extension on Lemma 5.1 in \cite{Cai_l1}, we know $\|A^Tz\|_\infty\leq \sigma\sqrt{(1+\delta_1^A)\log p}\leq\lambda/2$ with probability at least $1/\sqrt{\pi\log p}$. While for the matrix case, by Lemma 1.1 in \cite{Candes_Oracle}, we know $\|\mathcal{M}^\ast\|\leq 4\sigma\sqrt{\max(m,n)(1+\delta_1^A)\log 12}\leq \lambda/2$ with probability at least $1-e^{c\max(m,n)}$. Then in order to finish the proof, we only need to show \eqref{eq:signaloracle} or \eqref{eq:matrixoracle} given the assumption $\|A^Tz\|_\infty\leq \lambda/2$ or $\|\mathcal{M}^\ast(z)\|\leq\lambda/2$.
For the following part, we only give the proof for the signal case, since the matrix case is similar and the original proof by Cand\`es and Plan in \cite{Candes_Oracle} is already for the matrix case.
Define
$$K(\xi,\beta)=\gamma\|\xi\|_0+\|A\beta-A\xi\|_2^2,\quad \gamma=\frac{3\lambda^2}{16}=2\sigma^2\log p$$
Let $\bar{\beta}=\arg\min_{\xi}K(\xi,\beta)$, then we can deduce $\|\bar\beta\|_0\leq\|\beta\|_0\leq k$ by $K(\bar\beta,\beta)\leq K(\beta,\beta)$. By Lemma \ref{lm:delta2k<3deltak},
\begin{equation}\label{eq:hatbeta-barbeta}
\|\bar\beta-\beta\|_2^2\leq\frac{1}{1-\delta_{2k}^A}\|A\bar\beta-A\beta\|_2^2\leq\frac{1}{1-3\delta_k^A}\|A\bar\beta-A\beta\|_2^2
\end{equation}
By Lemma \ref{lm:barbeta}, we have
$$\|A^T(y-A\bar\beta)\|_\infty\leq\|A^T(y-A\beta)\|_\infty+\|A^TA(\beta-\bar\beta)\|_\infty\leq \lambda$$
which implies we can apply Theorem \ref{th:noisydantzigsignal} by plugging $\beta$ by $\bar\beta$:
$$\|\hat\beta-\bar\beta\|\leq\frac{\sqrt{2\|\bar\beta\|_0}}{1-3\delta_k^A}\cdot2\lambda$$
Hence,
\begin{eqnarray*}
\|\hat\beta-\beta\|_2^2&\leq&2\|\hat\beta-\bar\beta\|_2^2+2\|\bar\beta-\beta\|_2^2
  \leq \frac{16\|\bar\beta\|_0\lambda^2}{(1-3\delta_k^A)^2}+\frac{2}{1-3\delta_k^A}\|A\bar\beta-A\beta\|_2^2\\
 & \leq & \frac{256}{3(1-3\delta_k^A)^2}K(\bar\beta,\beta)
\end{eqnarray*}
Suppose $\beta'=\sum_{i=1}^p\beta_i1_{\{|\beta_i|>\mu\}}e_i$, where $e_i$ is the vector with 1 in the $i$th entry and 0 elsewhere, $\mu=\sqrt{\frac{\gamma}{1+\delta_k^A}}=\sqrt{\frac{3\lambda^2}{16(1+\delta_k^A)}}$. Then
\begin{eqnarray*}
K(\bar\beta,\beta)&\leq&K(\beta',\beta)
 \leq  \gamma\sum_{i=1}^p1_{\{|\beta_i|>\mu\}}+\|A\beta-A\beta'\|_2^2\\
 & \leq & \gamma\sum_{i=1}^p1_{\{|\beta_i|>\mu\}}+(1+\delta_k^A)\sum_{i=1}^p1_{\{|\beta_i|\leq \mu\}}|\beta_i|^2
 \leq  \sum_{i=1}^p\min\left(\gamma,(1+\delta_k^A)|\beta_i|^2\right)\\
 & \leq & 2\log p\sum_{i=1}^p\min(\sigma^2,|\beta_i|^2)
\end{eqnarray*}
The last inequality is due to $2\log p\geq (1+\delta_k^A)$. In summary, we get \eqref{eq:signaloracle}
given the assumption $\|A^Tz\|_\infty\leq \lambda/2$, which finishes the proof. \quad $\square$

\subsection{Technical Lemmas}
\label{lemmas.sec}

As seen in the proofs of Theorems \ref{th:mainsignal} and \ref{th:main}, it is necessary to estimate the left hand side of \eqref{eq:maininq1}, \eqref{eq:maininq2}, \eqref{eq:mainoddineq1}, \eqref{eq:mainoddineq2} and \eqref{eq:mainoddineq3}. Notice that these terms are of the similar type -- they are all the differences of the squared Frobenius norm of two matrices which only differ on a few leading terms in their SVD decompositions, we have the following lemma for the general estimation of this type of differences. Before we present the lemma, recall that we have defined the concept of indictor vector in Theorem \ref{th:mainsignal}.

\begin{Lemma}\label{lm:AX1-AX2}

For the vector case, suppose $g,h\geq 0, g+h\leq k$, $\{d_i\}_{i=1}^g, \{e_j\}_{j=1}^l, \{t_{ij}\}_{1\leq i\leq g,1\leq j\leq l}$ are non-negative real numbers satisfying
\begin{equation}
\min_{1\leq i\leq g} d_i \geq \max_{1\leq i\leq l} e_i,
\end{equation}
\begin{equation}
\sum_{i=1}^g t_{ij}= e_j,\quad \forall 1\leq j\leq l
\end{equation}
$\{b_i\}_{i=1}^h, \{c_i\}_{i=1}^h$ are real numbers. $\{u_{11},\cdots, u_{1h}; u_{31},\cdots, u_{3g}; u_{41},\cdots, u_{4l}\}$ is a set of indicator vectors with different support in $\mathbb{R}^p$; $\{u_{21},\cdots, u_{2h}; u_{31},\cdots, u_{3g}; u_{41},\cdots, u_{4l}\}$ is also a set of indicator vectors with different support. Define
$$\beta_1=\sum_{i=1}^hb_iu_{1i}+\sum_{i=1}^gd_iu_{3i}+\sum_{j=1}^le_ju_{4j}\in \mathbb{R}^{p}$$
$$\beta_2=\sum_{i=1}^hc_iu_{2i}+\sum_{i=1}^gd_iu_{3i}+\sum_{j=1}^le_ju_{4j}\in \mathbb{R}^{p}$$
Then we have
 \begin{equation}\label{eq:ineq1}
 \|A\beta_1\|_2^2-\|A\beta_2\|_2^2\geq (1-\delta_k^A)(\sum_{i=1}^hb_i^2+\sum_{i=1}^g(d_i+\sum_{j=1}^lt_{ij})^2)-(1+\delta_k^A)(\sum_{i=1}^hc_i^2+\sum_{i=1}^g(d_i+\sum_{j=1}^lt_{ij})^2)
 \end{equation}

For the matrix case, suppose $g,h\geq 0, g+h\leq r$, $\{d_i\}_{i=1}^g, \{e_j\}_{j=1}^l, \{t_{ij}\}_{1\leq i\leq g,1\leq j\leq l}$ are non-negative real numbers satisfying
\begin{equation}\label{eq:ineqd>e}
\min_{1\leq i\leq g} d_i \geq \max_{1\leq i\leq l} e_i,
\end{equation}
\begin{equation}\label{eq:tij=ei}
\sum_{i=1}^g t_{ij}= e_j,\quad \forall 1\leq j\leq l
\end{equation}
$\{b_i\}_{i=1}^h, \{c_i\}_{i=1}^h$ are real numbers. $\{u_{31},\cdots, u_{3g}; u_{41},\cdots, u_{4l}\}$ is a set of orthogonal unit vectors in $\mathbb{R}^m$, $\{u_{11},\cdots, u_{1h}\}$ and $\{u_{21},\cdots, u_{2h}\}$ are two sets of orthogonal unit vectors lying in the perpendicular space of $span\{u_{31},\cdots, u_{3g}; u_{41},\cdots, u_{4l}\}$; $\{v_{31},\cdots, v_{3g}; v_{41},\cdots, v_{4l}\}$ is a set of orthogonal unit vectors in $\mathbb{R}^n$, $\{v_{11},\cdots, v_{1h}\}$ and $\{v_{21},\cdots, v_{2h}\}$ are two sets of orthogonal unit vectors lying in the perpendicular space of $span\{v_{31},\cdots, v_{3g}; v_{41},\cdots, v_{4l}\}$. Define
$$X_1=\sum_{i=1}^hb_iu_{1i}v_{1i}^T+\sum_{i=1}^gd_iu_{3i}v_{3i}^T+\sum_{j=1}^le_ju_{4j}v_{4j}^T\in \mathbb{R}^{m\times n}$$
$$X_2=\sum_{i=1}^hc_iu_{2i}v_{2i}^T+\sum_{i=1}^gd_iu_{3i}v_{3i}^T+\sum_{j=1}^le_ju_{4j}v_{4j}^T\in \mathbb{R}^{m\times n}$$
Then we have
 \begin{equation}\label{eq:ineq1}
 \|\mathcal{M}(X_1)\|_2^2-\|\mathcal{M}(X_2)\|_2^2\geq (1-\delta_r^\mathcal{M})(\sum_{i=1}^hb_i^2+\sum_{i=1}^g(d_i+\sum_{j=1}^lt_{ij})^2)-(1+\delta_r^\mathcal{M})(\sum_{i=1}^hc_i^2+\sum_{i=1}^g(d_i+\sum_{j=1}^lt_{ij})^2)
 \end{equation}
\end{Lemma}

\noindent\textbf{Proof of Lemma \ref{lm:AX1-AX2}.} We prove the Lemma by induction on $l$.

When $l=0$, (\ref{eq:ineq1}) is clear to hold by the definition of $\delta_r^{\mathcal{M}}$ and the fact that $g+h\leq r$. Suppose (\ref{eq:ineq1}) holds for $l-1, (l\geq 1)$, we note
\begin{equation}
Y_i=-u_{3i}v_{3i}^T+u_{4l}v_{4l}^T,\quad 1\leq i\leq g
\end{equation}
\begin{equation}
P_z=X_z-\sum_{i=1}^gt_{il}Y_i,\quad z=1,2
\end{equation}
\begin{equation}
Q_{iz}=X_z-\sum_{w=1}^gt_{wl}Y_{w}+(t_{il}+d_i)Y_i\quad z=1,2,\quad 1\leq i\leq g
\end{equation}
We can show the following equality in $l_2$-space:
\begin{equation}\label{eq:AXz}
\begin{split}
&\mu\|\mathcal{M}(X_z-\sum_{i=1}^gt_{il}Y_i)\|_2^2+\sum_{i=1}^g\nu_i\|\mathcal{M}(X_z-\sum_{w=1}^gt_{wl}Y_{w}+(t_{il}+d_i)Y_i)\|_2^2\\
=&\|\mathcal{M}(X_z)\|_2^2+\mu\|\mathcal{M}(\sum_{i=1}^gt_{il}Y_i)\|^2_2+\sum_{i=1}^g\nu_i\|\mathcal{M}(-\sum_{w=1}^g t_{wl}Y_{w}+(t_{il}+d_i)Y_i)\|_2^2 \quad
\end{split}
\end{equation}
where $z=1,2$, $\nu_i=\frac{t_{il}}{d_i+t_{il}}$, $\mu=1-\sum_{i=1}^g\frac{t_{il}}{d_i+t_{il}}$. By (\ref{eq:ineqd>e}), (\ref{eq:tij=ei}) we have
$$\mu\geq 1-\sum_{i=1}^g\frac{t_{il}}{d_i}=1-\frac{e_l}{d_i}\geq 0$$
Thus, $\nu_i,\mu$ are all non-negative numbers satisfying $\mu+\sum_{i=1}^g \nu_i=1$. Consider the difference of these two equalities \eqref{eq:AXz} $(z=1,2)$, we get
\begin{equation}\label{eq:AX_1-AX_2}
\|\mathcal{M}(X_1)\|_2^2-\|\mathcal{M}(X_2)\|_2^2=\mu\left[\|\mathcal{M}(P_1)\|_2^2-\|\mathcal{M}(P_2)\|_2^2\right]+\sum_{i=1}^g\nu_i\left[\|\mathcal{M}(Q_{i1})\|_2^2-\|\mathcal{M}(Q_{i2})\|_2^2\right]
\end{equation}
By computing directly we can get
$$P_1=\sum_{i=1}^hb_iu_{1i}v_{1i}^T+\sum_{i=1}^g(d_i+t_{il})u_{3i}v_{3i}^T+\sum_{j=1}^{l-1}e_ju_{4j}v_{4j}^T$$
$$P_2=\sum_{i=1}^hc_iu_{2i}v_{2i}^T+\sum_{i=1}^g(d_i+t_{il})u_{3i}v_{3i}^T+\sum_{j=1}^{l-1}e_ju_{4j}v_{4j}^T$$
$$Q_{i1}=\sum_{w=1}^hb_wu_{1w}v_{1w}^T+\left[\sum_{w=1,w\neq i}^g(d_{w}+t_{wl})u_{3w}v_{3w}^T+(d_i+t_{il})u_{4l}v_{4l}^T\right]+\sum_{j=1}^{l-1}e_ju_{4j}v_{4j}^T$$
$$Q_{i2}=\sum_{w=1}^hc_wu_{2w}v_{2w}^T+\left[\sum_{w=1,w\neq i}^g(d_{w}+t_{wl})u_{3w}v_{3w}^T+(d_i+t_{il})u_{4l}v_{4l}^T\right]+\sum_{j=1}^{l-1}e_ju_{4j}v_{4j}^T$$
which corresponds with the assumption of $l-1$. Now by induction assumption of $l-1$, for all $1\leq w\leq g$ we have
\begin{equation}\label{eq:ineqPQ}
\begin{split}
&\|\mathcal{M}(P_1)\|_2^2-\|\mathcal{M}(P_2)\|_2^2\geq (1-\delta_r^\mathcal{M})(\sum_{i=1}^hb_i^2+\sum_{i=1}^g(d_i+\sum_{j=1}^lt_{ij})^2)-(1+\delta_r^\mathcal{M})(\sum_{i=1}^hc_i^2+\sum_{i=1}^g(d_i+\sum_{j=1}^lt_{ij})^2)\\
&\|\mathcal{M}(Q_{w1})\|_2^2-\|\mathcal{M}(Q_{w2})\|_2^2\geq (1-\delta_r^\mathcal{M})(\sum_{i=1}^hb_i^2+\sum_{i=1}^g(d_i+\sum_{j=1}^lt_{ij})^2)-(1+\delta_r^\mathcal{M})(\sum_{i=1}^hc_i^2+\sum_{i=1}^g(d_i+\sum_{j=1}^lt_{ij})^2)
\end{split}
\end{equation}
Together (\ref{eq:ineqPQ}) with (\ref{eq:AX_1-AX_2}), we can get (\ref{eq:ineq1}) for the case $l$.\quad$\square$

\begin{Lemma}\label{lm:ineq}
Suppose $m\geq r$, $a_1\geq a_2\geq\cdots\geq a_m\geq 0$, $\sum_{i=1}^ra_i\geq \sum_{i=r+1}^ma_i$, then for all $\alpha\geq1$,
\begin{equation}
\sum_{j=r+1}^ma_j^\alpha\leq \sum_{i=1}^r a_i^\alpha.
\end{equation}
More generally, suppose $a_1\geq a_2\geq \cdots\geq a_m\geq 0$, $\lambda\geq 0$ and $\sum_{i=1}^ra_i+\lambda\geq \sum_{i=r+1}^ma_i$, then for all $\alpha \geq 1$,
\begin{equation}\label{eq:ineqgeneral}
\sum_{j=r+1}^ma_j^\alpha\leq r\left(\sqrt[\alpha]{\frac{\sum_{i=1}^ra_i^\alpha}{r}}+\frac{\lambda}{r}\right)^\alpha
\end{equation}
\end{Lemma}

\noindent\textbf{Proof of Lemma \ref{lm:ineq}.}
It is sufficient to show the general part only. Since we can set $a_j=0$ when $j>m$, we assume $m\geq 2r$ without loss of generality. By Lemma \ref{lm:divide}, we can find $\{s_{ij}\}_{1\leq i\leq r,2r+1\leq j\leq m}$ satisfying
\eqref{eq:dividegeneral1}, \eqref{eq:dividegeneral2}. Hence,
\begin{eqnarray*}
\sum_{j=r+1}^ma_j^\alpha & = & \sum_{j=2r+1}^ma_j^{\alpha-1}(\sum_{i=1}^rs_{ij})+\sum_{j=r+1}^{2r}a_j^\alpha =  \sum_{i=1}^r\left(a_{r+i}^\alpha+\sum_{j=2r+1}^ma_j^{\alpha-1}s_{ij}\right)\\
 & \leq & \sum_{i=1}^r a_{r+i}^{\alpha-1}\left(a_{r+i}+\sum_{j=2r+1}^ms_{ij}\right)
 \leq  \sum_{i=1}^r\left(a_{r+i}+\sum_{j=2r+1}^ms_{ij}\right)^\alpha\\
 & \leq & r\left(\frac{\sum_{i=1}^ra_i}{r}+\frac{\lambda}{r}\right)^\alpha
  \leq  r\left(\sqrt[\alpha]{\frac{\sum_{i=1}^ra_i^\alpha}{r}}+\frac{\lambda}{r}\right)^\alpha.
  \quad \square
\end{eqnarray*}

\begin{Lemma}\label{lm:barbeta}
Suppose $\bar\beta=\arg\min_\xi K(\xi,\beta)$, then it satisfies
$\|A^TA(\bar\beta-\beta)\|\leq\lambda/2$.
\end{Lemma}

This is the vector version of Lemma 3.5 in \cite{Candes_Oracle}, for which we omit the proof here.


\end{document}